\documentclass[prd, aps, twocolumn, showpacs, nofootinbib,superscriptaddress]{revtex4}

\usepackage{amsmath,amssymb,amsthm,wasysym}
\usepackage{graphicx}
\usepackage{psfrag}
\usepackage{epstopdf}
\usepackage{color}
\usepackage{mathrsfs}
\usepackage{comment,ulem}

\def\lm{{\ell m}}

\def\v{v_\varphi}

\def\lm{{\ell m}}

\def\A{{\cal A}}

\def\F{{\cal F}}

\def\O{{\cal O}}

\def\N{{\cal N}}

\def\rhoo{S_\pm}
\def\rhoi{R_\pm}
\def\Msun{M_\odot}

\newcommand{\be}{\begin{equation}}
\newcommand{\ee}{\end{equation}}
\DeclareSymbolFontAlphabet{\mathrsfs}{rsfs}
\DeclareMathAlphabet{\mathcal}{OMS}{cmsy}{m}{n}
\newcommand{\scri}{{\mathrsfs{I}}}

\usepackage{float}

\definecolor{gray}{rgb}{.8,.8,.8}
\definecolor{cyan}{rgb}{0,0.9,0.9}
\definecolor{orange}{rgb}{0.9,0.5,0}
\definecolor{magenta}{rgb}{1,0,1}
\definecolor{purple}{rgb}{0.8,0.4,0.8}

\begin{document}

\title{Horizon-absorption effects in coalescing black-hole binaries: 
       An effective-one-body study of the non-spinning case.}

\author{Sebastiano \surname{Bernuzzi}}
\affiliation{Theoretical Physics Institute, University of Jena,
  07743 Jena, Germany}

\author{Alessandro \surname{Nagar}}
\affiliation{Institut des Hautes Etudes Scientifiques, 91440
  Bures-sur-Yvette, France}
                                   
\author{An{\i}l \surname{Zengino$\mathrm{\breve{g}}$lu}} 
\affiliation{Theoretical Astrophysics, California Institute of
  Technology, Pasadena, California, USA} 

\begin{abstract} 
We study the horizon absorption of gravitational waves in coalescing,
circularized, nonspinning black hole binaries. 
The horizon absorbed fluxes of a binary with a large mass ratio ($q=1000$)
obtained by numerical perturbative simulations are compared 
with an analytical, effective-one-body (EOB) resummed  expression
recently proposed. 
The perturbative method employs an analytical, linear in the mass
ratio, effective-one-body (EOB) resummed radiation reaction, and the
Regge-Wheeler-Zerilli (RWZ) formalism for wave extraction.
Hyperboloidal layers are employed for the numerical
solution of the RWZ equations to accurately compute horizon fluxes up to
the late plunge phase. The horizon fluxes from perturbative simulations 
and the EOB-resummed expression agree at the level of a few percent down 
to the late plunge. 
An upgrade of the EOB model for nonspinning binaries that includes  
horizon absorption of angular momentum as an additional term 
in the resummed radiation reaction is then discussed. 
The effect of this term on the waveform phasing for binaries
with mass ratios spanning $1$ to $1000$ is investigated.
We confirm that for comparable and intermediate-mass-ratio binaries
horizon absorbtion is practically negligible for detection 
with advanced LIGO and the Einstein Telescope (faithfulness $\geq 0.997$).
\end{abstract}


\maketitle

\section{Introduction}

The dynamics of the quasi-circular inspiral of coalescing binary black hole (BBH) 
systems is driven by the loss of mechanical angular momentum through gravitational 
radiation. The total loss of angular momentum consists of two
contributions: the one due to radiation emitted to future null infinity
($\F_\varphi^{\scri}$), and the one due to radiation absorbed by the 
black-hole horizons ($\F_\varphi^H$). 
Typically the former dominates over the latter, 
i.e.~$\F_\varphi^{\scri}\gg\F_\varphi^H$. For example, 
the leading order contribution to horizon absorption 
for a nonspinning binary is a 4PN contribution of
the form~\cite{Alvi:2001mx,Taylor:2008xy,Nagar:2011aa} 
\be
\label{eq:ratio}
\dfrac{\F^H}{\F^{\scri}_N} = x^4\left(1-4\nu + 2\nu^2\right)\left[1+c_1^H(\nu)x+\O(x^2)\right].
\ee
Above $x=(M\Omega)^{2/3}$ is the post-Newtonian orbital parameter, 
$\Omega$ is the orbital frequency, $M=M_A+M_B$ is the total mass of 
the system, with $M_{A,B}$ the masses of the individual black-holes, 
$\nu=M_A M_B/M^2$ is the symmetric mass ratio, and 
$\F^{\scri}_N=\nu^2\, 32/5\; x^{7/2}$ is the Newtonian contribution
to the asymptotic flux. The explicit expression of $c_1^H(\nu)$ 
follows from the state-of-the-art 1PN-accurate result of Taylor 
and Poisson~\cite{Taylor:2008xy}. 
In the presence of spin, a more complicated formula holds~\cite{Taylor:2008xy}, 
with the contribution of absorption entering already as a 2.5PN effect.
In practice, horizon absorption is a negligible effect when: 
(i) the separation between the two objects is large: (ii) the two objects have
comparable masses ($\nu\sim 1/4$); (iii) the spins are small.

Leading-order calculations by Alvi~\cite{Alvi:2001mx} 
(improved to 1PN fractional accuracy by Taylor and Poisson~\cite{Taylor:2008xy})
estimate the effect of horizon flows on the number of gravitational
wave (GW) cycles to be no more than $10\%$ of a cycle for
comparable-mass ($q=M_B/M_A=4$) binaries with maximally spinning
black holes by the time of merger (see Table~IV of
Ref.~\cite{Alvi:2001mx}). In the nonspinning case absorption effects
seem negligible with accumulated dephasings that are smaller than
$1\%$ of a cycle.  

The analysis of~\cite{Alvi:2001mx} is, however, inaccurate during the
late inspiral and plunge ($1/6\lesssim 
x \lesssim 1/3$). In this regime, absorption effects {\it may be
  relevant} for GW detection due to relativistic corrections, if the
mass ratio or the individual spins are sufficiently high. The
potential importance of absorption effects during the late plunge of
spinning binaries was also pointed out by Price and
Whelan~\cite{Price:2001un} using the close-limit approximation.

To meaningfully ascertain the importance of energy and angular momentum 
flows in or out of the black holes (depending on the orientation of the spin
with respect to angular momemtum) during the late
inspiral and plunge, one needs numerical relativity (NR) simulations.
The growth rate of the irreducible mass and angular momentum of the
black hole horizons in a NR simulation of nearly-extremal spinning
black hole binary~\cite{Lovelace:2011nu} has been compared to Alvi's
analytical prediction. A remarkable numerical agreement between the
two was found up to $x\lesssim 0.16$, while significant
deviations from numerical data were observed for larger
values of $x$. This result suggests that horizon-absorption effects
should be incorporated in the analytical modeling of coalescing black
hole binaries.

To bridge the gap between the leading-order estimate of Alvi 
valid during the early inspiral~\cite{Alvi:2001mx} 
and the qualitative understanding of Price and Whelan 
valid during the late plunge~\cite{Price:2001un} 
one needs an analytic description of the absorbed fluxes that
incorporates high-order PN corrections and that is not limited to the
slow-velocity, weak-field regime. 
Focusing on nonspinning binaries, Ref.~\cite{Nagar:2011aa} adapted
the resummation procedure of the asymptotic energy flux of 
Ref.~\cite{Damour:2008gu} to the energy flux absorbed by the
two black holes, so to consistently incorporate it within 
the effective-one-body (EOB)~\cite{Buonanno:1998gg,Buonanno:2000ef,Damour:2001tu} 
description of the dynamics of black hole binaries.
The final outcome  of that study is an analytical expression of 
the absorbed energy flux, written in a specific {\it factorized and resummed form}, 
that is well-behaved (contrary to a standard, PN expansion) 
also in the strong-field-fast-velocity regime (notably, also along 
the EOB-defined sequence of unstable circular orbits).
The input for the resummation procedure is given by state-of-the art
PN-expanded results for the horizon flux: the 1PN accurate expressions
of Taylor and Poisson~\cite{Taylor:2008xy} (valid for any mass ratio), 
and the leading-order results of Poisson and Sasaki~\cite{Poisson:1994yf}
in the test-mass ($\nu=0$) limit.
In addition, this analytical knowledge was further improved
by adding higher-order (effective) PN coefficients extracted
from the absorbed fluxes from circular orbits computed numerically
in the test-mass limit.
Finally, $\nu=0$ and $\nu\neq 0$ (semi)-analytical results
were hybridized to get improved accuracy for any mass ratio.

In this paper we study the effect of horizon-absorption on
the phasing of circularized, coalescing black-hole binaries 
up to merger. We do this by using the EOB description of the 
binary dynamics and radiation~\cite{Buonanno:1998gg,Buonanno:2000ef,Damour:2001tu}.
The radiation reaction is improved by an additional term, $\F_\varphi^H$, 
that takes into account the loss of mechanical angular momentum 
due to horizon absorption.  
As a first cut of the problem, we consider here {\it nonspinning} binaries
only, where the effects are weaker than when the BHs are 
spinning~\footnote{Note that the EOB approach can account
  consistently for (arbitrary) spins~\cite{Damour:2001tu,Barausse:2009xi,Barausse:2011ys,Nagar:2011fx,Taracchini:2012ig}. 
  Black hole absorption has already been included in EOB-based
  evolutions of extreme-mass-ratio (EMR) inspirals around a Kerr black
  hole, though only in its standard Taylor-expanded 
  form~\cite{Yunes:2010zj,Yunes:2009ef}. An improved treatment of this
  problem is currently under
  development~\cite{Taracchini:2012}.}.

First of all, we focus on the ``large-mass-ratio'' limit (e.g., $\nu=10^{-3}$) 
and we check the consistency of the ($\nu=0$) analytical expression of $\F_\varphi^H(x;\,0)$ 
proposed in~\cite{Nagar:2011aa} against the absorbed GW flux obtained 
numerically using a Regge-Wheeler-Zerilli (RWZ) perturbative treatment.
This gives further confirmation of the reliability of the resummation 
and hybridization procedure of the absorbed flux introduced in~\cite{Nagar:2011aa}.
Then we perform a comprehensive EOB study to investigate the 
effect of $\F_\varphi^H(x;\,\nu)$ on the phasing up to merger with
$10^{-3}\leq  \nu \leq 1/4$. Note that NR simulations for mass ratios
$q=100$ are currently doable~\cite{Gonzalez:2008bi,
Lousto:2010ut,Sperhake:2011ik},  
though they are challenging, do not yet provide sufficiently long
waveforms, and it does not seem practical to cover the parameter space
densely with full numerical relativity simulations only. 
Therefore, the EOB model is of fundamental importance to investigate
the so-called intermediate-mass-ratio (IMR)
regime~\cite{Brown:2006pj,Huerta:2010tp,Huerta:2011kt,Huerta:2011zi,Huerta:2012zy}.

The RWZ time-domain perturbative method employed in this work to obtain
large-mass-ratio waveforms is described 
in detail in~\cite{Nagar:2006xv,Damour:2007xr,Bernuzzi:2010ty,Bernuzzi:2010xj,Bernuzzi:2011aj}. 
We solve the RWZ equations for a binary system made of a point-particle 
on a Schwarzschild background and subject to leading-order, $\O(\nu)$, 
EOB-resummed analytical radiation reaction.
The main technical improvement introduced here is the development of 
smooth hyperboloidal layers~\cite{Zenginoglu:2010cq}
attached to a compact domain of Schwarzschild spacetime in standard
coordinates to include both future null infinity, $\scri$, and the 
black-hole horizon, $H$, in the computation. 
With this method, the absorbed and radiated fluxes can be computed 
very accurately. Also, the finite differencing order has been 
improved to 8th order accurate operators. These technical developments 
lead to such an efficient code that tail decay rates for the late-time
of the gravitational waveform emitted by inspiraling point particles 
can be computed accurately (this was not possible previously 
using standard methods).

This paper is organized as follows. In Sec.~\ref{sec:eob} we review the
results of Ref.~\cite{Nagar:2011aa} that are relevant for this work and
we give the explicit expression for $\F_\varphi^H$. 
In Sec.~\ref{sec:dlayer} we discuss the construction of hyperboloidal
layers and their advantages in improving the accuracy of the numerical
solution of the RWZ  
equation. In Appendix~\ref{sec:tails} we also demonstrate that the
layer technique helps solving a previously difficult problem of
obtaining accurate power law tails for inspiraling particles.
In Sec.~\ref{sec:smallnu} we present the RWZ calculation of the
absorbed waveforms and flux and the consistency check of
$\F^H_\varphi(x\;,0)$. The main results of the paper 
are collected in Sec.~\ref{sec:EOB_horizon}, where we investigate the
influence of $\F_\varphi^H$ on the phasing up to merger. Concluding
remarks are gathered in Sec.~\ref{sec:conc}. 
We use units with $G=c=1$.

\section{EOB dynamics and waveform: including horizon absorption}
\label{sec:eob}

In this Section we review the main elements of the EOB approach
and we recall the results of~\cite{Nagar:2011aa} that are needed 
to compute ${\cal F}_H$.
The EOB analytical description of the dynamics of a circularized binary 
essentially relies on two building blocks: the resummed EOB Hamiltonian 
$H_{\rm EOB}$, which describes conservative effects, and the 
resummed mechanical angular momentum loss $\F_\varphi$, 
which describes nonconservative effects due to loss of GW energy 
(radiation reaction)~\footnote{ %
  An additional radiation reaction term, $\F_r$, 
  is present  due to linear momentum loss through GWs, but, for
  circularized binaries, is typically not included because it remains
  small up to the late plunge.}.
The EOB Hamiltonian depends only on the relative position and momenta 
of the binary system. For nonspinning binaries it has the structure
\be
H_{\rm EOB}(r,p_{r_*},p_\varphi)\equiv M \sqrt{1+2\nu (\hat{H}_{\rm eff}-1)},
\ee
where
\be
\label{eq:Heff}
\hat{H}_{\rm eff}\equiv \sqrt{p_{r_*}^2+A(r)\left(1+
  \dfrac{p_\varphi^2}{r^2}+z_3\dfrac{p_{r_*}^4}{r^2}\right)}. 
\ee
Here $z_3\equiv 2\nu(4-3\nu)$ and we use rescaled dimensionless variables,
namely $r\equiv r_{AB} c^2/(G M)$, where $r_{AB} = |{\bf r}_A-{\bf r}_B|$, the relative 
separation between the two bodies, and $p_\varphi \equiv P_\varphi/(G M_A M_B)$, the angular momentum. 
In Eq.~\eqref{eq:Heff}, $p_{r_*}$ is the radial momentum canonically conjugate 
to a EOB-defined tortoise coordinate, $r_*$, that reduces to the usual tortoise 
coordinate when $\nu=0$. 
The function $A(r)$ is the basic radial potential that, following 
Ref.~\cite{Damour:2009kr},  depends on two EOB flexibility parameters 
$(a_5,\,a_6)$ that take into account effective 4PN and 5PN corrections 
to the conservative dynamics.
For coalescing black-hole binaries, an excellent phasing agreement
between NR and EOB waveforms can be reached over
banana-like regions in the $(a_5,a_6)$ plane.
Following Ref.~\cite{Damour:2009kr}, we fix the EOB parameters as
$a_5=-6.37$ and $a_6=50$ which lie within the extended region that yields 
a good fit with NR data for $q=1,\,2,$ and $4$. A recent 
study~\cite{Pan:2011gk} comparing an $(a_5,a_6)$-parametrized EOB 
model with NR simulations for $q=1,\,2,\,3,\,4,$ and $6$ (and more
accurate than those used in Ref.~\cite{Damour:2009kr}) pointed out that the 
``best fitting'' region in the $(a_5,a_6)$ plane actually depends on $\nu$
(see Fig.~5 in~\cite{Pan:2011gk}). Since our goal here is
to highlight only the effect of ${\cal F}_\varphi^H$ on the dynamics, 
we neglect this further $\nu$-dependence on $(a_5,a_6)$.
The analysis of the $\nu$-dependence of $(a_5,a_6)$ in the calibration of the
EOB model of Ref.~\cite{Damour:2009kr}, in the presence of black-hole absorption
and with better numerical data, is postponed to future work.

The radiation reaction force, $\F_\varphi$, drives the
angular momentum loss during evolution.  The Hamilton equation 
for $p_\varphi$ reads
\be
\dfrac{dp_\varphi}{dt} = \hat{\F}_\varphi,
\ee
where $\hat{\F}_\varphi=\F_\varphi/\nu$. 
The mechanical angular momentum loss is typically written as
\be
\label{eq:Fphi}
\hat{\F}_\varphi = -\dfrac{32}{5} \nu r_\omega^4 \Omega^5 \hat{f}(\v^2;\,\nu).
\ee 
Here, $\Omega=d\varphi/dt$ is the orbital frequency,
with $\varphi$ the orbital phase, 
$\v=r_\omega \Omega$ is the azimuthal velocity,
and $r_\omega=r\psi^{1/3}$, where 
$\psi$ is a $\nu$-dependent correction factor that is necessary  
to formally preserve Kepler's law during the plunge~\cite{Damour:2006tr}.  
The function $\hat{f}(x;\,\nu)$ is the {\it reduced flux function}
that is defined, for a circularized binary, as the ratio between the
total energy flux and the $\ell=m=2$ asymptotic energy flux. In our
case the reduced flux function is given by the sum of an asymptotic
and a horizon contribution as 
\be
\label{eq:fhat}
\hat{f}(x;\,\nu) = \hat{f}^\scri(x;\,\nu) + \hat{f}^H(x;\,\nu),
\ee
where each term  is given by
\begin{align}
\hat{f}^{(\scri,H)}(x;\,\nu) &= F^{(\scri,H)}_{\ell_{\rm
    max}}/F_{22}^{\rm N}. 
\end{align}
Here, $F^{(\scri,H)}_{\ell_{\rm max}}$ are the total asymptotic
($\scri$) and horizon ($H$) energy fluxes for circular orbits summed
up to multipole $\ell=\ell_{\rm max}$,  while $F_{22}^N = (32/5) \nu^2
x^5$ is the Newtonian quadrupolar (asymptotic) energy flux.  
In the EOB model one uses suitably factorized expressions for the
multipolar fluxes $F^{(\scri,H)}_{\lm}$ to resum and improve them
with respect to standard PN-expanded expressions in the strong-field,
fast-velocity regime ($1/6\lesssim x\lesssim 1/3$). The resummation of
the asymptotic waveform and fluxes was discussed in
Ref.~\cite{Damour:2008gu} and has been used in many works since
then. We use it here at the $3^{+2}$PN accuracy~\footnote{ %
  The 3PN-accurate  $\nu$-dependent terms are augmented by the 4PN and 
  5PN~accurate $\nu=0$  corrections for all multipoles.}
and we fix $\ell_{\rm max}=8$. 

The horizon flux is written as the sum (up to $\ell_{\rm max}=8$) 
\be
F^{H,(\ell_{\rm max})}(x;\,\nu) = \sum_{\ell =2}^{\ell_{\rm
    max}}\sum_{m=1}^{\ell} F_\lm^{(H,\epsilon)}(x;\,\nu) 
\ee
where the partial multipolar fluxes have the following factorized
structure~\cite{Nagar:2011aa} 
\be
\label{eq:FlmH}
F_\lm^{(H,\epsilon)}(x;\,\nu)=F_{\lm}^{(H_{\rm
    LO},\epsilon)}(x;\,\nu)\left[\hat{S}_{\rm eff}^{(\epsilon)}(x;\nu)
  \left(\rho_\lm^H(x;\,\nu)\right)^\ell\right]^2. 
\ee
Here, $\epsilon\equiv \pi(\ell+m)=0,1$ is the parity of the considered
multipole,  $\hat{S}_{\rm eff}^{(\epsilon)}$ is a source factor, with
$\hat{S}_{\rm eff}^{(0)}=\hat{H}_{\rm eff}$ or $\hat{S}_{\rm
  eff}^{(1)}=\sqrt{x}p_\varphi$ according to the parity of the
multipole, and the $\rho_\lm^H(x;\,\nu)$ are the residual amplitude
corrections to the horizon waveform. Only $\rho_{22}^H(x;\,\nu)$ is
known analytically at 1PN accuracy~\cite{Nagar:2011aa}. It reads 
\be
\label{eq:rho22_1PN}
\rho_{22}^{H_{\rm 1PN}}(x;\,\nu)=1+\dfrac{4-21\nu+27\nu^2-8\nu^3}{4(1-4\nu+2\nu^2)}x + \O(x^2).
\ee
To improve our knowledge of the strong-field behavior of the $\rho_{22}^H$ functions, 
Ref.~\cite{Nagar:2011aa} computed numerically the $\rho_\lm^{H_{\rm num}}$ functions
for a test particle moving on (stable and unstable) circular orbits on a Schwarzschild
background. For each multipole, it was possible to fit the numerically computed 
$\rho_\lm^{H_{\rm num}}$ accurately via a suitable  rational function of the form
\be
\label{eq:rhofit}
\rho_\lm^{H_{\rm fit}}(x)=\dfrac{1+n_1^\lm x + n_2^\lm x^2 + n_3^\lm x^3 +
  n_4^\lm x^4}{1+d_1^\lm x + d_2^\lm x^2} 
\ee
where $n_i^\lm$ and $d_i^\lm$ are free fitting
parameters~\footnote{ %
  Note that for the $\ell=m=2$ mode the fit was done
  imposing the constraint that the 1PN coefficient is equal to 1, 
  because $\rho_{22}^H(x;\,0) = 1 + x + \O(x^2)$.}. 
By Taylor-expanding Eq.~\eqref{eq:rhofit} in powers of $x$ one obtains
the following representation of the $\rho_\lm^H$ functions in the
$\nu=0$ limit 
\be
\label{eq:rholm_taylor}
\rho_\lm^H(x;\,0) = T_N[\rho_\lm^{H_{\rm fit}}(x)],
\ee
where $N$ indicates the maximum power of the expansion. For the
$\ell=m=2$ mode, Ref.~\cite{Nagar:2011aa} pointed out that  setting
$N=4$ (i.e., 4PN accuracy) is sufficient to yield an accurate
representation of the $\rho_\lm^{H_{\rm num}}$ up to and below the
last-stable-orbit (LSO) at $r=6$, with relatively small 
differences around the light-ring
(see Fig.~3 of~\cite{Nagar:2011aa}). We have verified that this
remains true also for the other multipoles, so that we shall assume
4PN accuracy in Eq.~\eqref{eq:rholm_taylor} from now on. Following
Ref.~\cite{Nagar:2011aa}, we {\it hybridize} the $\nu$-dependent 1PN
information of Eq.~\eqref{eq:rho22_1PN} with the 4PN expansion of
Eq.~\eqref{eq:rholm_taylor}. Such hybridization procedure, that is
conceptually analogous to what has been done in 
Ref.~\cite{Damour:2008gu} for
the corresponding asymptotic residual amplitude corrections, is
justified in view of the following two results of
Ref.~\cite{Nagar:2011aa}: (i) the dependence on $\nu$ of the 1PN
coefficient in  Eq.~\eqref{eq:rho22_1PN} is mild; (ii) the fit of the
numerical data proved to be robust enough so that the coefficients of
the PN expansion can be taken as reliable estimates for the actual
(yet un-calculated) PN coefficients. In practice we use the following
4PN expression for the $\rho_\lm^H(x;\,\nu)$ 
\be
\label{eq:all_rho}
\rho_\lm^H(x;\,\nu) = 1 + c^\lm_1 x +  c^\lm_2 x^2 +  c^\lm_3 x^3 + c^\lm_4 x^4 .
\ee
The values of the coefficients $c^\lm_i$, $i=1,\dots,4$ are listed in
Table~\ref{tab:clm}, where in fact only $c_1^{22}$ is given
analytically as a function of $\nu$, while the other coefficients are
computed from the test-mass $n^\lm_i$ and $d^\lm_i$  coefficients
extracted from the fit. We shall use them in the following as an
effective representation of the  actual test-mass information,
although the hope is that it will be soon possible to replace them
with terms from a PN calculation.

\begin{table}[t]
  \caption{\label{tab:clm} Coefficients of our hybrid
    1$^{+3}$PN-accurate $\rho_\lm^H(x;\,\nu)$ functions as given by
    Eq.~\eqref{eq:all_rho}.} 
  \begin{center}
    \begin{ruledtabular}
      \begin{tabular}{cccccc}
        $\ell$   &   $m$  &    $c_1^\lm$   &   $c_2^\lm$  &   $c_3^\lm$   &   $c_4^\lm$ \\
        \hline 
        2    &    2   &  $\frac{4-21\nu + 27\nu^2
          -8\nu^3}{4(1-4\nu+2\nu^2)}$   &  4.78752  & 26.760136  &
        43.861478  \\ 
        2    &    1   &   0.58121   &  1.01059  & 7.955729    & 1.650228 \\
        3    &    3   &   1.13649   &  3.84104  & 45.696716   & 27.55066  \\
        3    &    2   &   0.83711   &  1.39699  &  23.638062  & -1.491898 \\
        3    &    1   &   1.61064   &  2.97176  &  10.045280  &  15.146875 \\
        4    &    4   &   1.15290   &  4.59627  &  55.268737  & 13.255971 \\
        4    &    3   &   0.96063   &  1.45472  & 43.480636   &-35.225828  \\
        4    &    2   &   1.43458   &  2.43232  & 21.927986   & 10.419841\\
        4    &    1   &   0.90588   &  1.17477  & 5.126480    & 4.022307 
   \end{tabular}
  \end{ruledtabular}
\end{center}
\end{table}

In Table~\ref{tab:clm} we list all PN coefficients up to $\ell=4$.
It seems enough to include only the quadrupolar
contributions $\rho_{21}^H$ and $\rho_{22}^H$ in $\hat{f}^H(x;\,\nu)$,
since, as we show in Sec.~\ref{sec:smallnu} below, the effect of
multipoles with $\ell\geq 3$ on the horizon-absorbed angular-momentum
flux is practically negligible already in small-mass-ratio coalescence
events.
[Note that the $\nu$-dependence of the leading-order
  prefactor to the multipolar horizon flux,  $F_\lm^{H_{\rm LO}}$ is
  fully known only for the quadrupole modes~\cite{Taylor:2008xy}].  

Using Eqs.~\eqref{eq:fhat}, \eqref{eq:FlmH} and \eqref{eq:all_rho} one
defines an EOB dynamics that takes into account
horizon absorption. From this dynamics one then computes the
(asymptotic) EOB multipolar waveform that has the well known
factorized structure 
\be
\label{eq:hlm}
h_\lm = h_\lm^{(N,\epsilon)} \hat{S}^{(\epsilon)}_{\rm eff}\hat{h}^{\rm tail}_\lm (\rho_\lm)^\ell\hat{h}^{\rm NQC}_\lm,
\ee
where $h_\lm^{(N,\epsilon)}$ is the Newtonian waveform, $\hat{h}^{\rm
  tail}\equiv T_\lm e^{i\delta_\lm}$ is the tail factor as defined in
Ref.~\cite{Damour:2008gu}, $\rho_\lm$ is the resummed modulus
correction and $\hat{h}^{\rm NQC}_\lm$ is a next-to-quasi-circular
correction. For each multipole $(\ell,m)$ these NQC corrections depend
on 4 parameters, $a_i^{\ell m}$, $i=1,\dots,4$ 
(two for amplitude corrections and two for a phase correction) 
that have to be determined with an iterative procedure 
to match the EOB waveform to the NR waveform around merger. 
The NQC correction to the amplitude 
depending on $(a_1^\lm,a_2^\lm)$ is the same as in 
Refs.~\cite{Damour:2009kr,Bernuzzi:2010xj}; the NQC correction 
to the phase depending on $(a_3^\lm,a_4^\lm)$ is implemented as 
per Eq.~(22) of Ref.~\cite{Pan:2011gk}, that proved more robust 
than the analogous expression used in Eq.~(12) of~\cite{Bernuzzi:2010xj}
to complete the EOB waveform in the extreme 
mass-ratio limit. The $a_i^\lm$ parameters are determined as
in~\cite{Bernuzzi:2010xj} by imposing that the slope of the EOB
waveform amplitude and frequency agree with the
NR ones at the peak of the EOB orbital frequency $\Omega$. 
Note that, consistently with the findings 
of~\cite{Bernuzzi:2010xj} and differently from previous 
work~\cite{Damour:2009kr,Pan:2011gk}, we {\it do not} impose 
that the peak of $|h_{22}|$ occurs at the same time as the peak
of $\Omega$. On the contrary, we allow $|h_{22}|$ to have a nonzero 
slope there that coincides with the slope of the NR waveform modulus
$|h_{22}^{\rm NR}|$ at a NR time that occurs slightly after the time 
corresponding to $\max|h_{22}|$. This NR-data extraction point is 
suitably chosen consistently with the test-mass results~\cite{Damour:2012prep}.
To obtain the coefficients $a_i^{\lm}$ for any value of $\nu$, 
we fit with cubic polynomials in $\nu$ the NR points 
extracted from both the waveforms computed for us by D.~Pollney and C.~Reisswig  
using the Llama code~\cite{Pollney:2009yz,Damour:2011fu,Damour:2012prep}, 
for mass ratios $q=1,2,3,4$, and the perturbative
data of~\cite{Bernuzzi:2010xj,Bernuzzi:2011aj}.
As a last step we match to the EOB inspiral-plus-plunge waveform,
Eq.~\eqref{eq:hlm}, a superposition of Kerr black-hole quasi-normal-modes
(QNMs) over a matching ``comb''~\cite{Damour:2007xr}. We use in
general five QNMs; note, however, that for $\nu=0$ three QNMs 
are sufficient to obtain good agreement between EOB 
and RWZ waveforms~\cite{Bernuzzi:2010xj}.

\section{Transmitting layers for the Regge-Wheeler-Zerilli equations}
\label{sec:dlayer}

In this Section we describe the hyperboloidal layers adopted
here to solve the RWZ equations and to extract the GW fluxes at the
horizon and at null infinity. The method
builds on previous
work~\cite{Zenginoglu:2007jw,Zenginoglu:2010cq,Bernuzzi:2011aj} and
extends the hyperboloidal layer technique to the near-horizon regime.
We also present, as a test of the implementation, the horizon absorbed
fluxes from geodesic circular motion, and, in
Appendix~\ref{sec:tails}, we report tail computations with our
new infrastructure. 

\subsection{Smooth hyperboloidal layers}
We use the Schwarzschild time coordinate $t$ and the tortoise
coordinate $r_*$ in the bulk for describing the inspiralling
particle using the standard EOB formalism. The tortoise coordinate  
\be\label{eq:tort} r_* = r + 2 M \log (r-2M),\ee
is constructed such that the event horizon $r=2M$ is at infinite
coordinate distance. From a numerical point of view, the main effect
of the tortoise coordinate is to push away the coordinate singularity at
the bifurcation sphere in Schwarzschild coordinates. 
The computational domain is then truncated at some negative value 
for $r_*$ and ingoing boundary conditions are applied.  

There are two problems with this common approach. 
First, the artificial truncation of the computational domain leads to 
artificial boundary conditions. This problem is not as important in the 
negative $r_*$ direction as in the positive one, because the potential falls
off exponentially in the tortoise coordinate towards the horizon 
whereas only polynomially towards spatial infinity. Nevertheless, the imposition of 
such artificial boundary conditions can still complicate the implementation
of higher order discretization methods. 
Second, the computation of absorbed fluxes 
by the black hole is performed at finite radius. 
To avoid contamination of the horizon flux computation by the artificial boundary
conditions, a large grid in the negative $r_*$ direction needs to be chosen
(see, for example, \cite{Martel:2003jj}). 
This practice leads to a waste of computational resources.

A resolution to these problems is to change the coordinates near the horizon and in the asymptotic domain ("near infinity"), 
while keeping the standard Schwarzschild coordinates in the bulk. In our previous
studies \cite{Bernuzzi:2010xj, Bernuzzi:2011aj} we applied 
hyperboloidal scri-fixing in a layer \cite{Zenginoglu:2007jw, Zenginoglu:2010cq} 
to solve these problems near infinity.
In its original form, such a hyperboloidal layer is attached in the positive 
radial direction only so that the outer boundary corresponds to future null infinity.
Since we are using the tortoise coordinate $r_*$, a similar layer can 
be attached also in the negative $r_*$ direction so that the inner boundary corresponds
to the black hole horizon. The time foliation in this layer is then not hyperboloidal 
but horizon penetrating. Nevertheless, we will keep using the term hyperboloidal layer
for this new construction because the foliation has hyperboloidal properties in the
tortoise coordinate.

The method consists of a spatial coordinate compactification 
and a time transformation as described below.  

\subsubsection{Spatial compactification.}
Consider a finite domain $\mathcal{D}$ in the tortoise coordinate
$r_*$ given by $\mathcal{D}=[-R_-,R_+]$ where $R_\pm\in
\mathbb{R}^+$. In this finite domain, we use coordinates $(t,r_*)$.  
We introduce a compactifying coordinate\footnote{For notational continuity
with previous work we use the same symbol to address both the compactifying 
coordinate and the residual amplitude corrections $\rho_\lm$ to the EOB waveform.} 
$\rho$ to calculate the solution to the RWZ equations numerically 
on the unbounded domains $(-\infty, -R_-)$ and $(R_+,\infty)$. 
The compactification is such that the infinities are mapped to a 
finite $\rho$, and at the interfaces $R_\pm$ the coordinates $\rho$ and $r_*$ agree. 

A convenient way to write such a compactification is 
\be
\label{eq:space} r_\ast = \frac{\rho}{\Omega(\rho)},
\ee
where $\Omega(\rho)$ is a suitable function of $\rho$. It is unity in
the bulk domain, $\Omega_{\mathcal{D}}=1$, implying
$\rho=r_*$ on $\mathcal{D}$. For compactification, $\Omega$ must
vanish at a finite $\rho$ location, which then corresponds to infinity
with respect to $r_*$ (see \cite{Zenginoglu:2010cq, Bernuzzi:2011aj}
for details). The transformation therefore is degenerate at the zero
set of $\Omega$. Its Jacobian reads
\be
J\equiv \dfrac{d\rho}{dr_*}=\dfrac{\Omega^2}{\Omega-\rho\,\Omega'},
\ee
where the prime indicates $d/d\rho$. 
A simple prescription for $\Omega$ to compactify both directions could be
\be\label{eq:c4_om} \Omega = 1- \left(\frac{|\rho| - R_\pm}{S_\pm - R_\pm}\right)^4
\Theta(|\rho|- R_\pm)\,,\ee
For $\rho<0$ we use the plus sign, for $\rho>0$ we use the minus sign in
the above formula. The transformation \eqref{eq:space} with \eqref{eq:c4_om} 
maps the unbounded domain $-\infty<r_*<+\infty$ to
the bounded domain $-S_-<\rho<S_+$ such that $\rho=r_*$ on
$\mathcal{D}=[-R_-,R_+]$ where $S_\pm>R_\pm$.

The choice of $\Omega$ in \eqref{eq:c4_om} leads to a coordinate
transformation that is $C^4$ at the interfaces. Our numerical
experiments showed that this degree of smoothness was not sufficient for the
accurate computation of late-time tail decay rates of the waveform reported in 
Appendix~\ref{sec:tails}. 
Numerical studies of hyperboloidal compactification using RWZ equations 
previously showed that a smooth ($C^\infty$) transition leads to 
higher accuracy~\cite{Zenginoglu:2009ey}. 
Such a smooth transition function can be given as 
\begin{eqnarray*} 
\label{eq:f_t} f_T := \frac{1}{2}+\frac{1}{2}\tanh \left[
\frac{s}{\pi} \left( \tan x- \frac{q^2}{\tan x} \right) \right],
\end{eqnarray*}
where we have defined
\[ x:= \left(\frac{\pi}{2} \frac{|\rho|-\rhoi}{\rhoo-\rhoi}\right). \]
The free parameter $q$ determines the point $\rho_{1/2}$ at which
$f_T(\rho_{1/2})=1/2$ and $s$ determines the slope of $f_T$ at
$\rho_{1/2}$ \cite{Yunes:2005nn, Vega:2009qb}. We set 
\be \label{eq:smooth_om} \Omega = 1 - \frac{|\rho|}{S_\pm} f_T\,
\Theta(|\rho| - R_\pm).\ee

\begin{figure}[t]
\center
\includegraphics[width=0.37\textwidth]{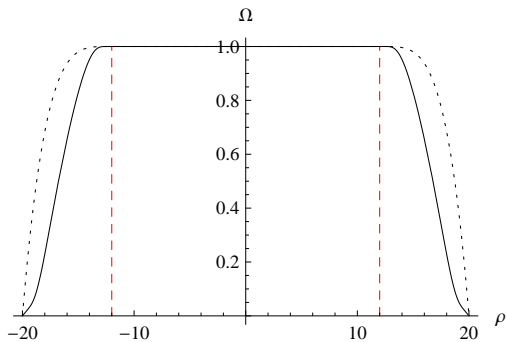}
\caption{\label{fig:compress}
The function $\Omega$ for the two choices \eqref{eq:c4_om} and
\eqref{eq:smooth_om}. The dashed vertical lines (red online) indicate the interfaces at
$R_\pm=12$. Infinity corresponds to $S_\pm= 20$. The
dashed (black online) curve denotes the $C^4$ transition~\eqref{eq:c4_om}, 
and the solid (black online) curve the smooth transition~\eqref{eq:smooth_om}.
The numerical results obtained later in the text are obtained with the smooth
transition and with this choice of parameters.}
\end{figure}

The two choices for $\Omega$ have been plotted in Fig.~\ref{fig:compress}. 
For the main numerical results in this paper we use the smooth 
compactification of Eq.~\eqref{eq:smooth_om}.  

\subsubsection{Time transformation.}
It is well known that spatial compactification alone leads to
resolution problems for hyperbolic equations~\cite{Orszag}. The loss
of resolution near infinity, however, can be avoided for essentially outgoing
solutions by combining the spatial compactification with a suitable time
transformation~\cite{Zenginoglu:2010cq}. The details of this
transformation depend on the background spacetime, but the 
essential idea is to keep the outgoing null direction invariant 
in local compactifying coordinates~\cite{Bernuzzi:2011aj}. 

A suitable time transformation for numerical computations keeps the
background metric invariant of the time coordinate by respecting the
timelike Killing field~\cite{Zenginoglu:2007jw}. Such time
transformations can be written in the following form 
\be
\label{eq:tau}
\tau = t\pm h(r_*) \ ,  
\ee
where the function $h$ is called the {\it height function} and
depends on the tortoise coordinate only.  

Near the black hole horizon, and near null infinity, gravitational
waves propagate predominantly in one direction along null rays. Near
the black hole most waves are absorbed, near infinity most waves
escape. Correspondingly, near the black hole we require invariance of
ingoing null rays in local coordinates, whereas near infinity we
require invariance of outgoing null rays. The sign in Eq.~\eqref{eq:tau}
depends therefore on the sign of $r_*$. The invariance of the null
direction in local compactifying coordinates translates into 
\[ t\pm r_* = \tau \pm \rho .\]
With Eq.~\eqref{eq:tau} we get
\[ r_* = \rho + h \]
or by defining $H:=dh(r_*)/dr_*$
\be\label{eq:rel}  H  = 1 - J . \ee
This relation between the differential time transformation $H$ and the
differential spatial compactification $J$ solves the resolution problem of 
compactification in hyperbolic equations. 

We emphasize that, even though the inner hyperboloidal layer 
changes the time foliation, we do not modify the particle 
trajectory consistently when solving the RWZ equation. In principle, the particle motion
should be expressed in the local coordinates of the inner layer. 
In practice, however, this seems unnecessary when the layer 
is attached at a sufficiently small negative value of $r_*=-R_-<0$. 
We find that after the particle has crossed the light ring 
at $3M$, thereby triggering the QNM ringdown, its subsequent trajectory does not 
influence the waveform. Choosing $R_-=12$ allows us to leave 
the description of the particle untouched.
Once the particle enters the layer, we smoothly switch off the RWZ source 
to avoid unphysical features in the ringdown waveform 
(see Fig.~16 of~\cite{Bernuzzi:2011aj}).

\subsection{Horizon fluxes for circular orbits}

\begin{figure}[t]
\center
\includegraphics[width=0.5\textwidth]{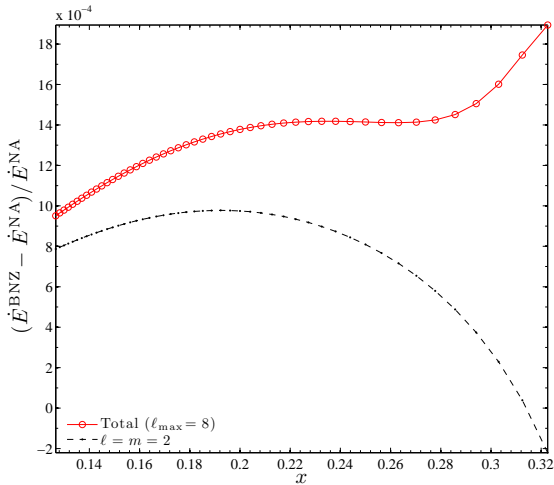}
\caption{\label{fig:ciro} (color online)
  Testing the accuracy of the updated time-domain RWZ code
  for a particle along a sequence of stable and unstable circular orbits.
  We plot the fractional difference in the horizon fluxes computed with the
  time-domain RWZ code using hyperboloidal layers and 
  S.~Akcay's frequency-domain code~\cite{Akcay:2010dx,Nagar:2011aa}.}
\end{figure}

As a test of the accuracy of our new numerical setup, and in particular, 
of the inner layer, we consider a point-particle moving on 
circular orbits of a Schwarzschild black hole and we compute the
horizon fluxes. The treatment of the distributional $\delta$-function
describing the point-particle source as a finite-size, narrow Gaussian
is the same as previous works~\cite{Nagar:2006xv,Damour:2007xr,Bernuzzi:2010ty,Bernuzzi:2010xj,Bernuzzi:2011aj}.   
Given a selected sample of stable and unstable orbits of radius $r$ 
($3.1\leq r \leq 7.9$ spaced by $\Delta r=0.1$), the RWZ waveform 
at the horizon location
$\Psi_{\lm}^{(H,\epsilon)}$, and its time derivative,
$\dot{\Psi}_{\lm}^{(H,\epsilon)}$,  the fluxes of energy and angular momentum 
absorbed by the black hole are given by~\cite{Martel:2003jj}
\begin{align}
\label{eq:Edot}
\dot{E}^H_{(\ell_{\rm max})} &=\dfrac{1}{16\pi}\sum_{\ell =2}^{\ell_{\rm max}}\sum_{m=0}^{\ell}\sum_{\epsilon=0}^1
\dfrac{(\ell+2)!}{(\ell-2)!}|\dot{\Psi}^{(H,\epsilon)}_\lm|^2\\ 
\label{eq:Jdot}
\dot{J}^H_{(\ell_{\rm max})} &=-\dfrac{1}{8\pi}\sum_{\ell =2}^{\ell_{\rm max}}\sum_{m=1}^{\ell}\sum_{\epsilon=0}^1 m
\dfrac{(\ell+2)!}{(\ell-2)!}\Im\left[\dot{\Psi}^{(H,\epsilon)}_\lm
  \Psi^{(H,\epsilon)*}_\lm \right] \ . 
\end{align}
In Fig.~\ref{fig:ciro} we show the fractional difference 
(plotted versus $x=1/r=(M\Omega)^{2/3}$) between the the energy 
flux $\dot{E}^H$ computed with our code (labeled by ``BNZ'') 
and the same quantity obtained by S.~Akcay using his frequency 
domain code~\cite{Akcay:2010dx}, and presented for the first 
time in Ref.~\cite{Nagar:2011aa} (labeled by ``NA''), i.e.,
$(\dot{E}^{\rm BNZ}-\dot{E}^{\rm NA})/\dot{E}^{\rm NA}$.
The solid (red online) curve in the plot refers to the total flux 
summed up to $\ell_{\rm max}=8$, while the dashed one to the $\ell=m=2$ 
dominant quadrupole mode only.
The frequency domain computation of horizon fluxes using
the code of Ref.~\cite{Akcay:2010dx} have fractional 
uncertainty of order $10^{-10}$ or smaller for strong-field orbits 
(say $r\leq 10$). Figure~\ref{fig:ciro} highlights how
the fractional difference between the fluxes obtained with the two
methods is on the order of $10^{-3}$.

\section{Horizon absorption in the large-mass-ratio limit}
\label{sec:smallnu}

\subsection{Perturbative, time-domain computation}
In this section we compute the horizon-absorbed GW fluxes in a
large-mass-ratio BBH coalescence using the perturbative method
discussed extensively in previous 
works~\cite{Nagar:2006xv,Damour:2007xr,Bernuzzi:2010ty,Bernuzzi:2010xj,Bernuzzi:2011aj}.   
The computations allow us to test the reliability of the
EOB-resummed fluxes given by Eq.~\eqref{eq:FlmH}.

\begin{figure}[t]
\center
\includegraphics[width=0.5\textwidth]{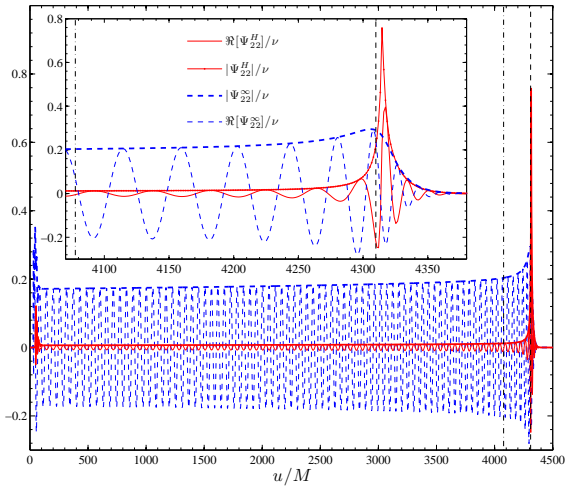}
\caption{\label{fig:insp:waves} (color online)
  Comparing horizon and null infinity quadrupolar ($\ell=m=2$) RWZ
  waveforms for a coalescing binary with mass ratio $\nu=10^{-3}$.  The
  horizontal axis corresponds to horizon anticipated time $u\equiv
  u^+= \tau+H$ for the horizon waveform and to null infinity retarded
  time, $u\equiv u^-=\tau-S$ for the asymptotic waveform. The leftmost
  (dash-dotted) vertical line marks the (dynamical) time when the particle crosses
  the LSO, while the rightmost (dashed)  vertical line corresponds 
  instead to the light-ring crossing. The horizon waveform (red online) 
  becomes unreliable around the light-ring crossing ($u/M\gtrsim 4300$). 
  See text for discussion.} 
\end{figure}

In the large-mass-ratio limit the EOB Hamiltonian tends to the
Schwarzschild one, and  higher-order corrections 
in the analytical radiation reaction are neglected. The radiation-reaction term
is then given by
\be
\label{eq:RR_test_mass}
\hat{\F}_\varphi \equiv \hat{\F}_\varphi^\scri+\hat{\F}_\varphi^H=
-\dfrac{32}{5}\nu
r^4\Omega^5\left[\hat{f}^\scri(\v^2;\,0)+\hat{f}^H(\v^2;\,0)\right], 
\ee
with $\v=r\Omega$. Here, $\hat{f}^{\scri}(\v^2;\,0)$ is computed as in 
Ref.~\cite{Damour:2008gu} in the $\nu=0$ limit
but retaining all terms up to 5PN fractional accuracy in the
$\rho_\lm$'s computed in Ref.~\cite{Fujita:2010xj} (see also Ref.~\cite{Fujita:2011zk} 
for the 14PN accurate calculation).

We work here with the mass ratio~\footnote{ %
  Note that in the test-mass limit, $M_A/M_B\ll 1$ we can 
  identify the inverse mass ratio $1/q=M_A/M_B$ with the {\it symmetric} 
  mass ratio $\nu=M_A M_B/(M_A + M_B)^2$.}
$\nu=10^{-3}$. Previous studies~\cite{Bernuzzi:2010ty,Bernuzzi:2011aj} 
indicated that, in this case, the method gives a fractional agreement 
between the 5PN-accurate mechanical angular momentum loss and the actual 
angular momentum flux computed from the RWZ master function of order $10^{-3}$
even beyond the LSO (see Fig.~14 of~\cite{Bernuzzi:2011aj}.) The RWZ
master function is extracted numerically using the method of
Sec.~\ref{sec:dlayer}. Neglecting horizon absorption 
in the dynamics ($\hat{f}^H(\v^2;\,0)=0$, in Eq.~\eqref{eq:RR_test_mass}), 
we reproduce the relative dynamics of previous 
works~\cite{Bernuzzi:2010ty,Bernuzzi:2010xj,Bernuzzi:2011aj}.
The initial relative separation is $r_0=7$ and the relative 
dynamics is started with the usual post-circular initial 
conditions~\cite{Buonanno:2000ef,Nagar:2006xv}. 

Figure~\ref{fig:insp:waves} focuses on the $\ell=m=2$ mode and
illustrates the relative importance of the horizon waveform
$\Psi^{(H,0)}_{22}$ compared  to the asymptotic  
waveform $\Psi^{(\scri,0)}_{22}$. The figure shows on the 
same panel the real part of the waveforms together with their amplitudes.
In the strong--field regime under consideration, $r\lesssim 7$, the horizon 
waveform is smaller ($\sim 16$ times during inspiral) than the asymptotic 
waveform but not negligible 
(roughly comparable to some asymptotic subdominant multipoles).
Notably, one finds that $|\Psi^{(H,0)}_{22}|$ is always {\it larger} than  
$|\Psi^{(\scri,0)}_{44}|$. The ratio between the two varies between 1.5 
at the beginning of the inspiral up to 2 at LSO crossing.

\begin{figure}[t]
\center
\includegraphics[width=0.5\textwidth]{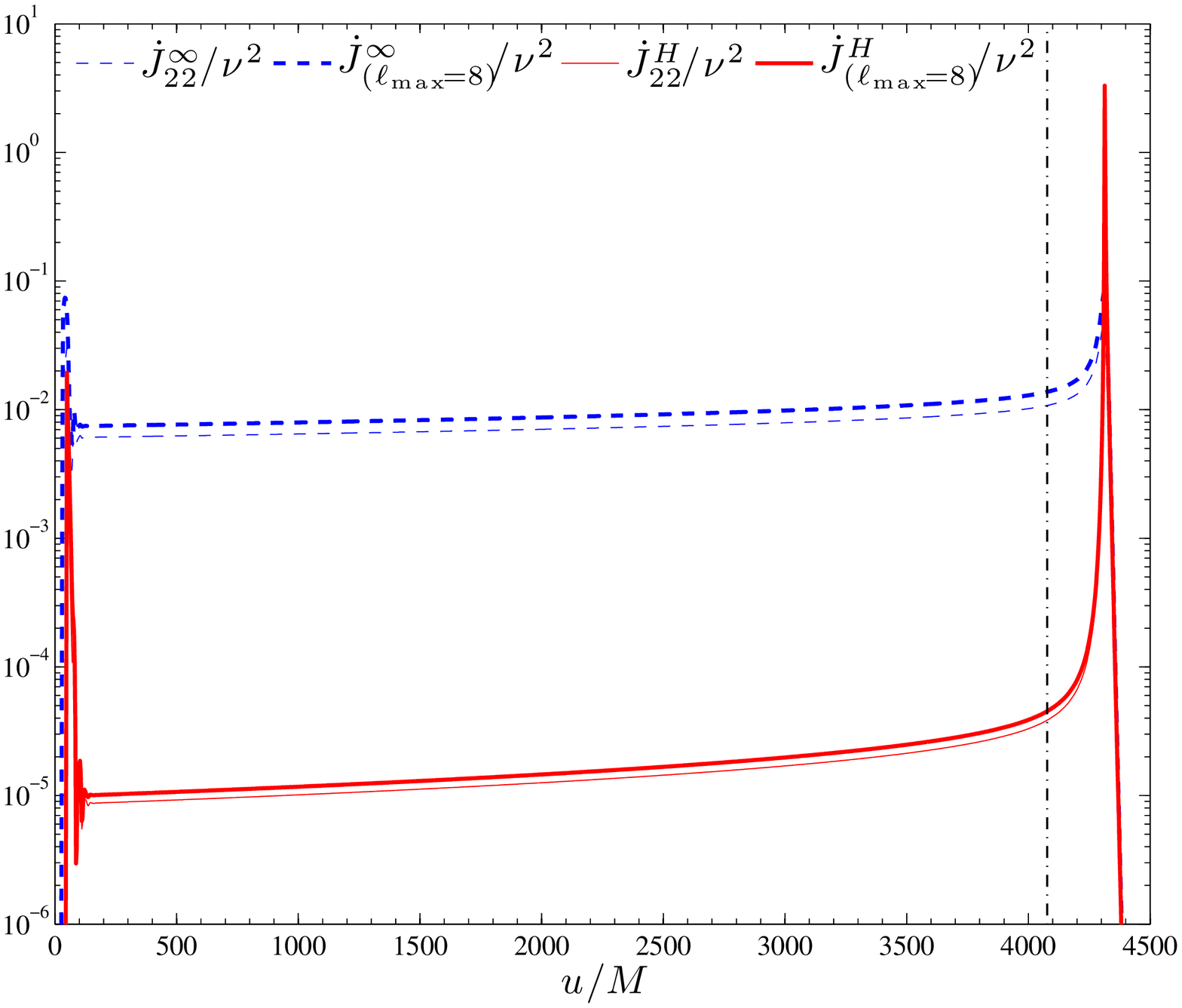}\\
\includegraphics[width=0.5\textwidth]{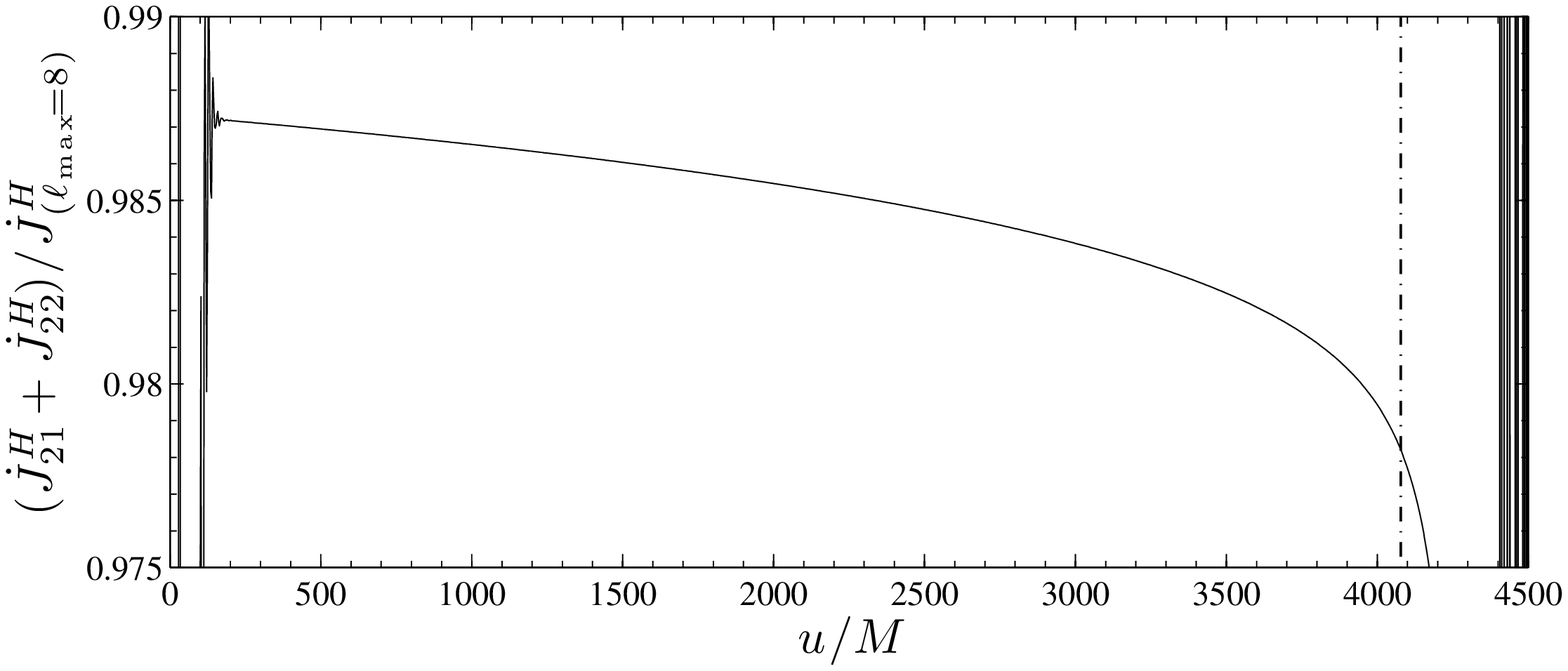}
\caption{\label{fig:insp:jfluxs} (color online).
  Top panel: comparison between RWZ horizon and asymptotic angular
  momentum fluxes for mass ratio $\nu=10^{-3}$ from Eq.~\eqref{eq:Jdot} with 
  $\ell_{\rm max}=8$. Bottom panel: the $\ell=2$ modes contribute to more than 
  the $98\%$ of the total absorbed flux up to LSO crossing (vertical dashed line).}  
\end{figure}

\begin{figure*}[t]
\center
\includegraphics[width=0.49\textwidth]{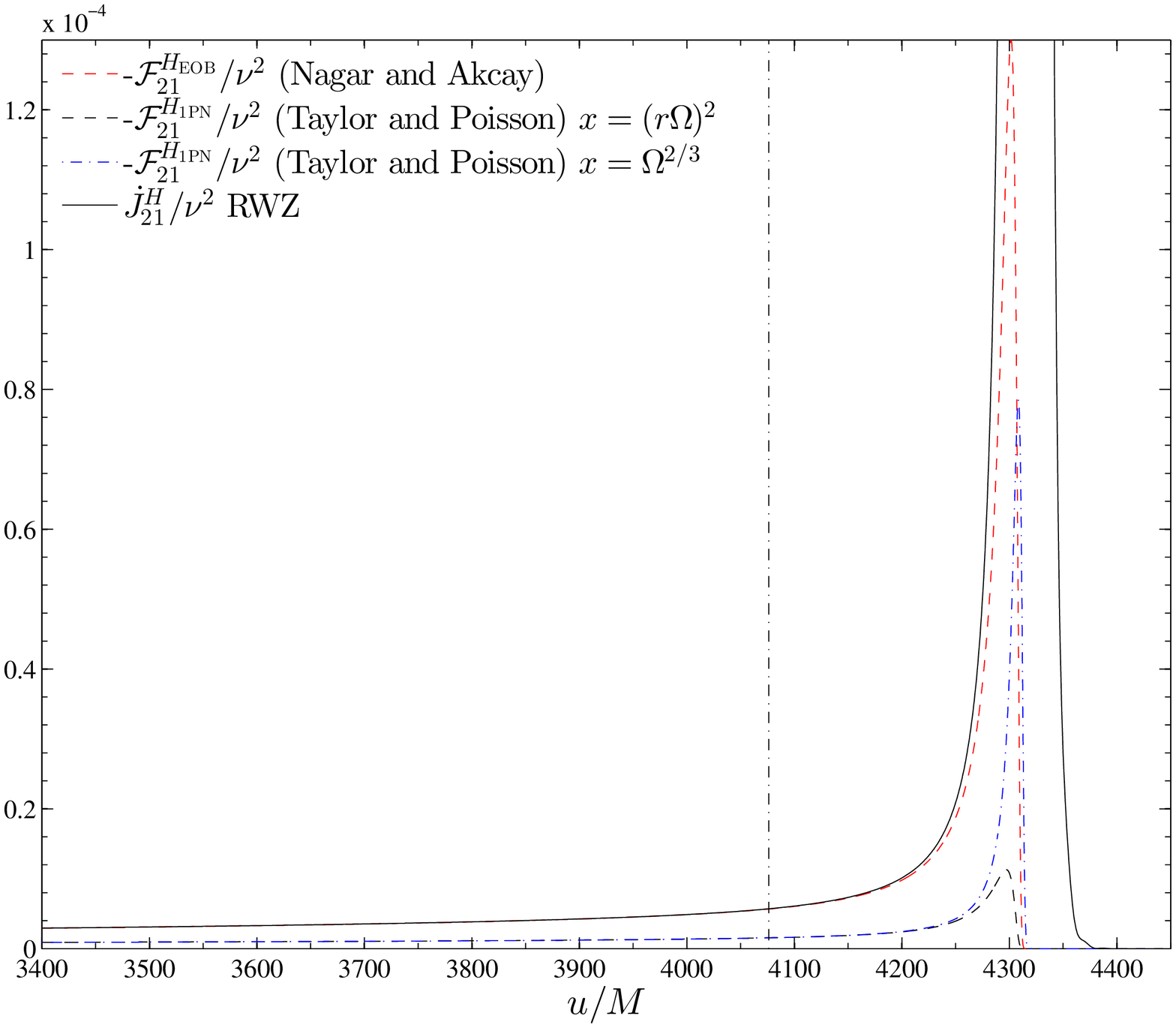}
\includegraphics[width=0.49\textwidth]{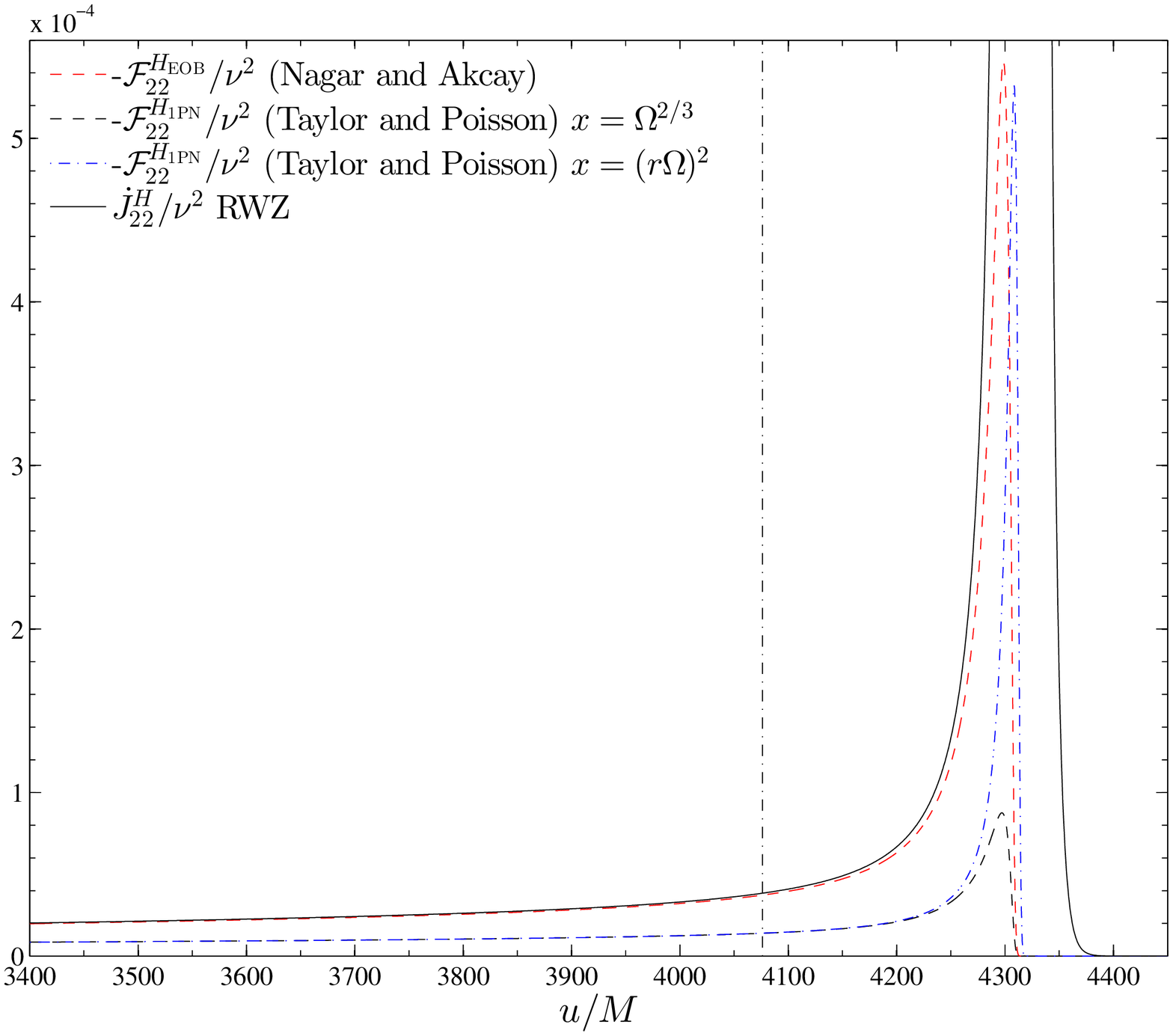}
\caption{\label{fig:probe}(color online) Comparison between EOB resummed angular momentum flux and the RWZ one
for the quadrupole ($\ell=2$) modes: $m=1$ (left panel) and $m=2$ (right panel).
The dash-dotted vertical line line marks the LSO crossing. The EOB-resummed (horizon) mechanical
angular momentum loss shows very good consistency with the horizon flux computed from GWs. By contrast,
the 1PN-accurate expressions, Eqs.~\eqref{eq:TP22}-\eqref{eq:TP21}, underestimate horizon absorption 
by more than a factor 2.} 
\end{figure*}

The amplitude of the horizon waveform grows during the late plunge and
reaches about  $0.1$ just before the light-ring crossing, $u/M\approx 4300$.
It then increases by a factor 
$\sim 7$ over a temporal interval $\sim 15$, developing a ``spike'' that is twice as large as the
corresponding value of the asymptotic amplitude. After this transient, 
the ringdown asymptotic and horizon waveforms are consistent.

The presence of a spike in the horizon waveform is due to our representation
of the point-particle source as a narrow ($\sigma\ll 1$) Gaussian.
The RWZ function is (in the $\sigma\to 0$ limit) discontinuous at $r_*=R(t)$ and its spatial
derivative is singular. Since we have not implemented a sophisticated 
regularization of the source 
(see in this respect Refs. ~\cite{Lousto:1996sx,Martel:2001yf,Barack:2009ux,Poisson:2011nh}),
there is a spatial (smoothed) singularity on the RWZ computational grid at 
the particle location. 
After the particle has crossed the light ring, the singularity is advected to the
horizon. The presence of such a discontinuity in the RWZ function and
the corresponding singularity in the energy flux (also observed 
in the analytical treatment of an extreme-mass-ratio plunge by Hamerly and Chen~\cite{Hamerly:2010cr}),
makes our numerical representation of the particle ill-suited for a detailed study of horizon 
absorption during the last moments of the merger. We have, however, verified that
the effect is localized around the location of the particle and its influence is reduced for smaller values of $\sigma$.
In this work, we use the RWZ horizon
waveform (and flux) only before the light-ring crossing, say  $u/M\sim 4300$, 
so that our results are not affected by the absorption of the particle by the horizon.

We display in  Fig.~\ref{fig:insp:jfluxs} the horizon-absorbed angular momentum flux $\dot{J}^H_{\ell_{\rm max}}/\nu^2$
computed from Eq.~\eqref{eq:Jdot} with $\ell_{\rm max}=8$.
The top panel contrasts asymptotic fluxes (either summed up to $\ell_{\rm max}=8$ or just $\ell=m=2$),
with the horizon fluxes, highlighting that the latter are typically $10^{-3}$ times smaller.
The bottom panel of the figure shows the ratio between the total quadrupole horizon flux
(i.e., $\dot{J}^H_{21}+\dot{J}^H_{22}$) and the total horizon flux  $\dot{J}^H_{(\ell_{\rm max}=8)}$, 
which indicates that the quadrupole mode accounts for more than the $98\%$ of the absorption 
up to the LSO crossing (dash-dotted vertical line in the plot).

\subsection{The EOB-resummed horizon flux}
\label{sec:smallnu:eob}

We compare the horizon absorbed angular momentum flux computed from the RWZ waveform,
Eq.~\eqref{eq:Jdot}, with the EOB-defined mechanical angular momentum loss 
due to horizon absorption, Eq.~\eqref{eq:RR_test_mass}.
In this section, the dynamics is computed including only 
$\hat{\F}_\varphi^\scri$; the effect of $\F_\varphi^H$ is explored in
the next section.
Figure~\ref{fig:probe} shows the dominant quadrupole $\ell=2$ fluxes for $m=1$ 
(left panel) and $m=2$ (right panel).
The mechanical losses $-\hat{\F}^H_{22}/\nu$ computed with various approximations
(non-solid lines) are contrasted with $\dot{J}_{2m}^H/\nu^2$ (solid lines)
The vertical dash-dotted line marks the LSO crossing.
In addition to the EOB resummed analytical expressions (dashed curves, red online),
we also show the PN-expanded (1PN-accurate) absorbed fluxes as
computed by Taylor and Poisson~\cite{Taylor:2008xy},
(see also Eq.~(13) of~\cite{Nagar:2011aa}). They are given by
\begin{align}
\label{eq:TP22}
-\F_{22}^{H_{\rm 1PN}}(x)& = \dfrac{32}{5}\nu^2 x^{15/2}(1+3x),\\
\label{eq:TP21}
-\F_{21}^{H_{\rm 1PN}}(x)& = \dfrac{32}{5}\nu^2 x^{17/2}.
\end{align}
When plotting these expressions we use two different PN representations of $x$:
either $x\equiv v_\phi^2$ (dashed-line, black online) consistently with 
the EOB waveform,  or $x_\Omega = \Omega^{2/3}$ (dash-dotted line, blue online).
The two expressions differ only well below the LSO due to the violation of the
Kepler constraint during the plunge.

Following observations can be made in Fig.~\ref{fig:probe}. First, the PN expanded expressions 
clearly underestimate the absorbed flux in the strong-field regime. 
This is expected due to the structure of the $\rho_\lm^H$ in the circular case. It has been shown 
in Ref.~\cite{Nagar:2011aa} (Fig.~3) that at $x=1/7\approx 0.14$ the 1PN-accurate $\rho_{22}^H$ is more than 
a factor of two smaller than the corresponding $\rho_{22}^{H_{\rm num}}$ computed from numerical data. 

Second, the EOB resummed expression (with the {\it fitted} coefficients $c_i^{\lm}$) 
shows a very good consistency with the exact angular momentum flux computed from the waves. 
For the $\ell=m=2$ mode, the fractional difference is~$\approx 1\%$ at the beginning of the 
inspiral, to grow then up to~$\approx 3\%$ at the LSO crossing. 
Notably, an excellent agreement occurs also for the $m=1$ flux
(fractional difference $<1\%$ at LSO crossing), where the knowledge of 
the function $\rho_{21}^H$ comes completely from the fit to the circular data~\cite{Nagar:2011aa}.
The fractional difference we find here is approximately one order of magnitude larger than for the
asymptotic flux (for the same mass ratio $\nu=10^{-3}$), see Fig.~14 of~\cite{Bernuzzi:2011aj}. 
This difference is not surprising because we have little analytical information to compute the EOB horizon flux.
The computation relies mostly on the coefficients $c_i^\lm$ obtained from the fit 
to the numerical data. 

Third, the fluxes stay close also {\it below} the LSO crossing, 
even though we do not expect the RWZ fluxes to be accurate close to the light-ring crossing. 
The fact that the fluxes remain so close during the late inspiral up to the plunge is by itself a confirmation 
that the fitted $c_i$'s yield a rather accurate approximation to the coefficients one would get 
from the analytic PN calculation. 

In conclusion we have shown that the analytical expression of $\F^H_\varphi$, 
built using several pieces of information coming from a circularized binary
(either analytical or numerical) shows an excellent agreement with the
exact horizon flux computed from the RWZ waves. This makes us
confident that we can safely use $\F^H_\varphi$ as a new term in the
radiation reaction to take horizon absorption into account. The
influence of this term on the waveform phasing will be discussed in
detail below.

\subsection{Effect on BBH phasing}
\label{sec:EOB_horizon}

\begin{figure}[t]
\center
\includegraphics[width=0.5\textwidth]{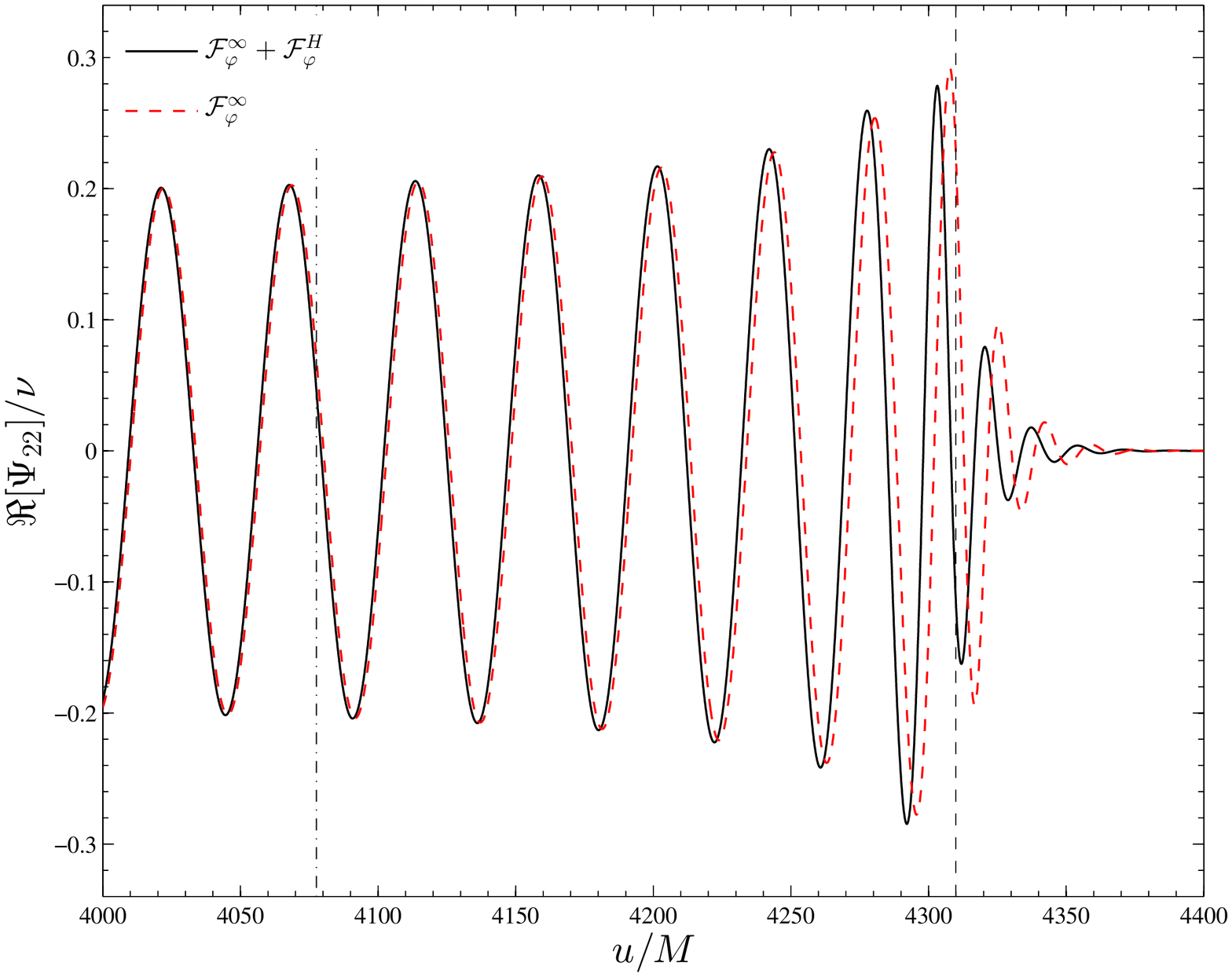}\\
\includegraphics[width=0.5\textwidth]{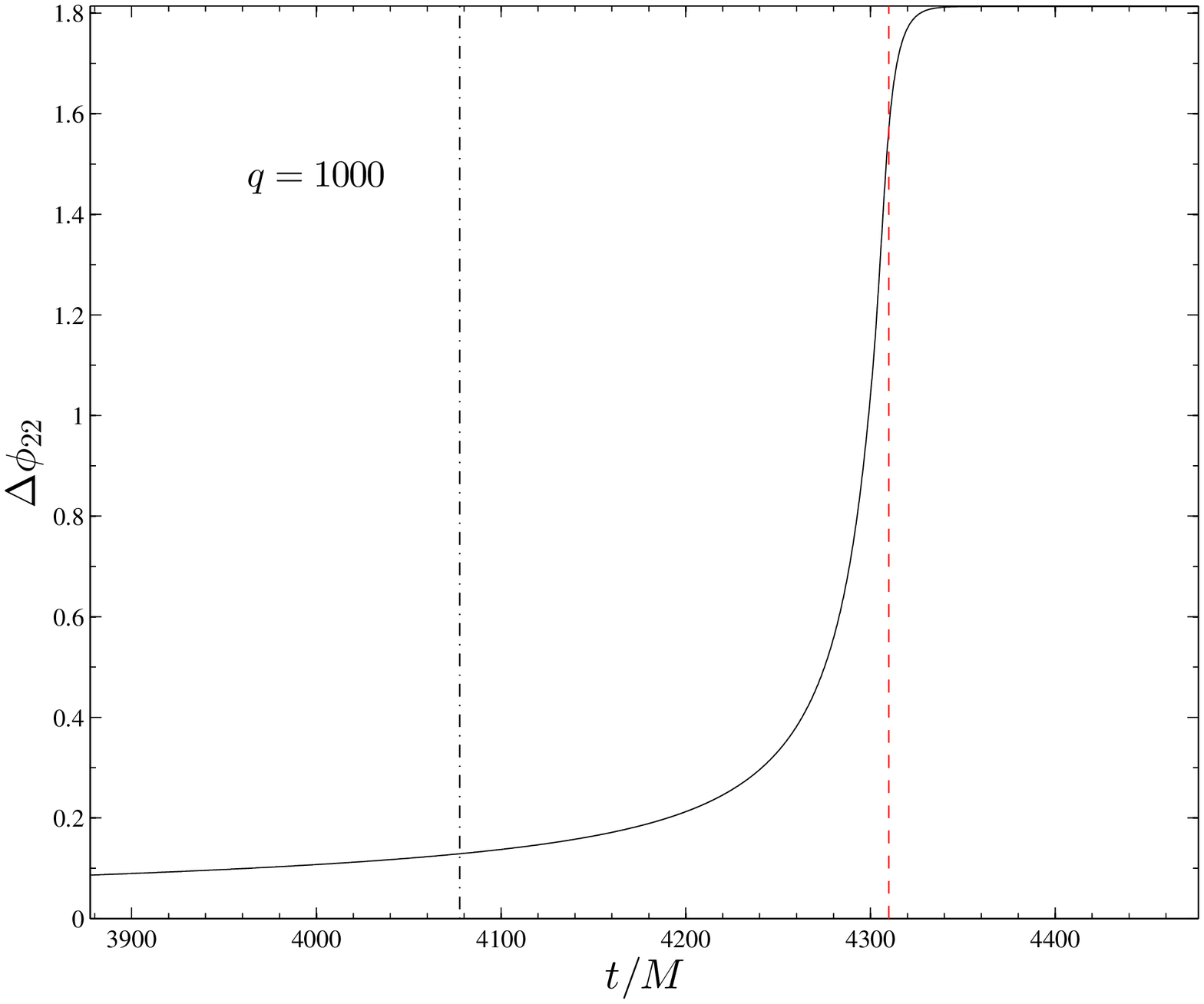}
\caption{\label{fig:dphi_test_mass} (color online) Test-mass limit ($\nu=10^{-3}$): including $\F_\varphi^H(v\,;0)$ in the dynamics
and its effect on the $\ell=m=2$ phasing.  The top panel compares the $\ell=m=2$ EOB waveforms with (solid line) and without (dashed line) 
$\F_\varphi^H(v;\,0)$.  The accumulated phase difference (bottom panel) is of order 0.1~rad at LSO crossing (dash-dotted vertical line), 
and reaches a remarkable 1.5~rad at merger (dashed vertical line).}
\end{figure}

\begin{figure*}[t]
\center
\includegraphics[width=0.49\textwidth]{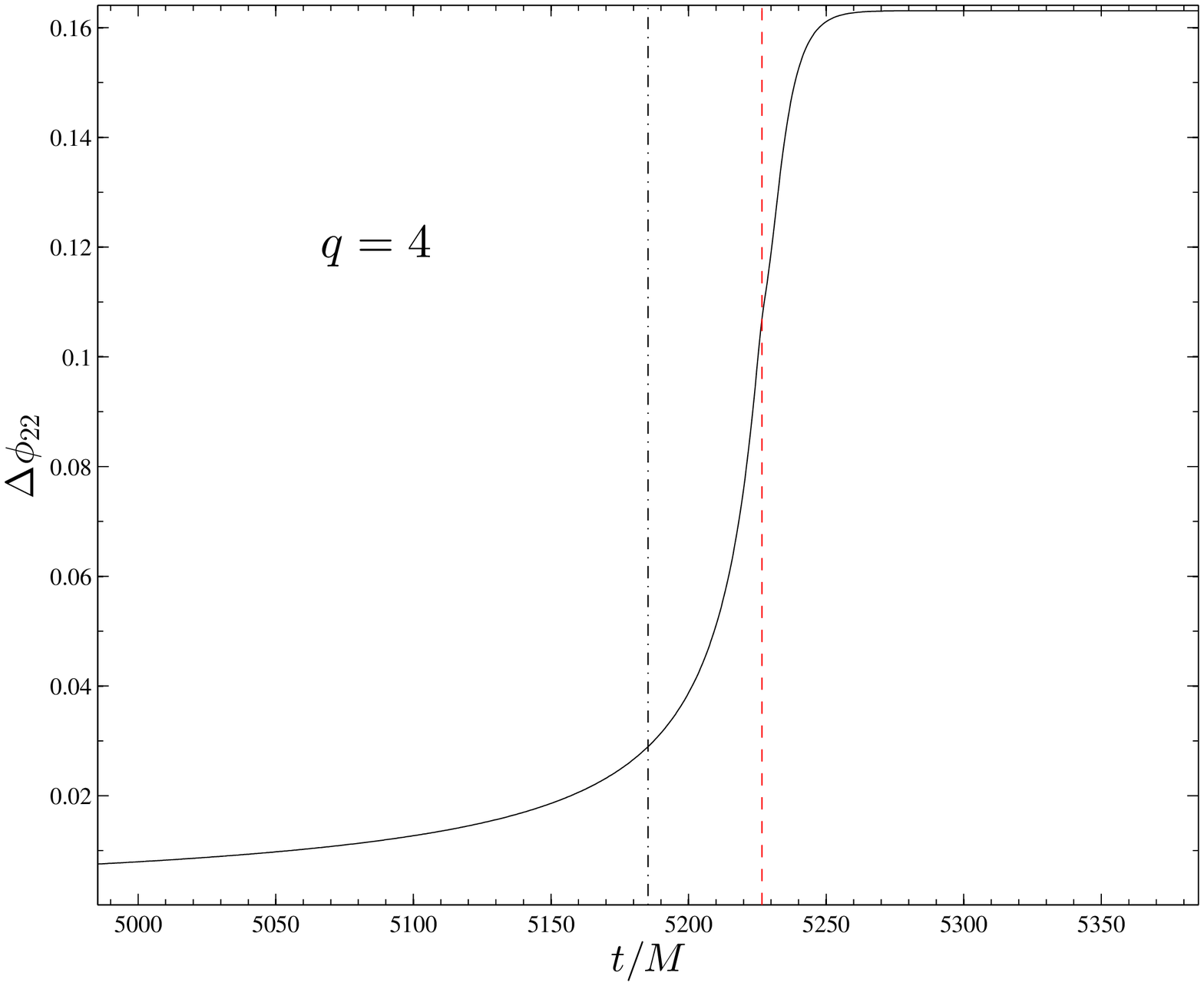}
\includegraphics[width=0.49\textwidth]{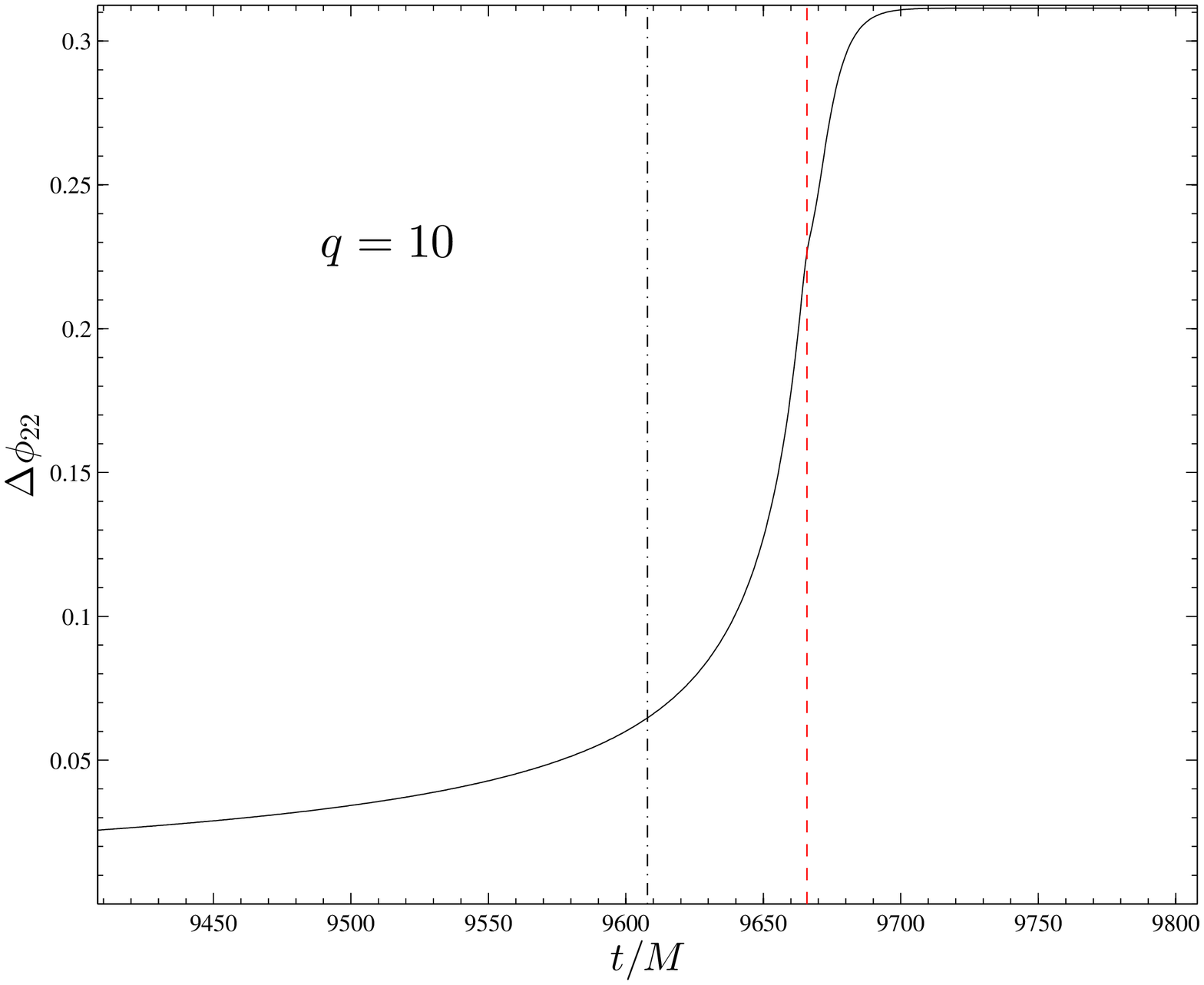}\\
\includegraphics[width=0.49\textwidth]{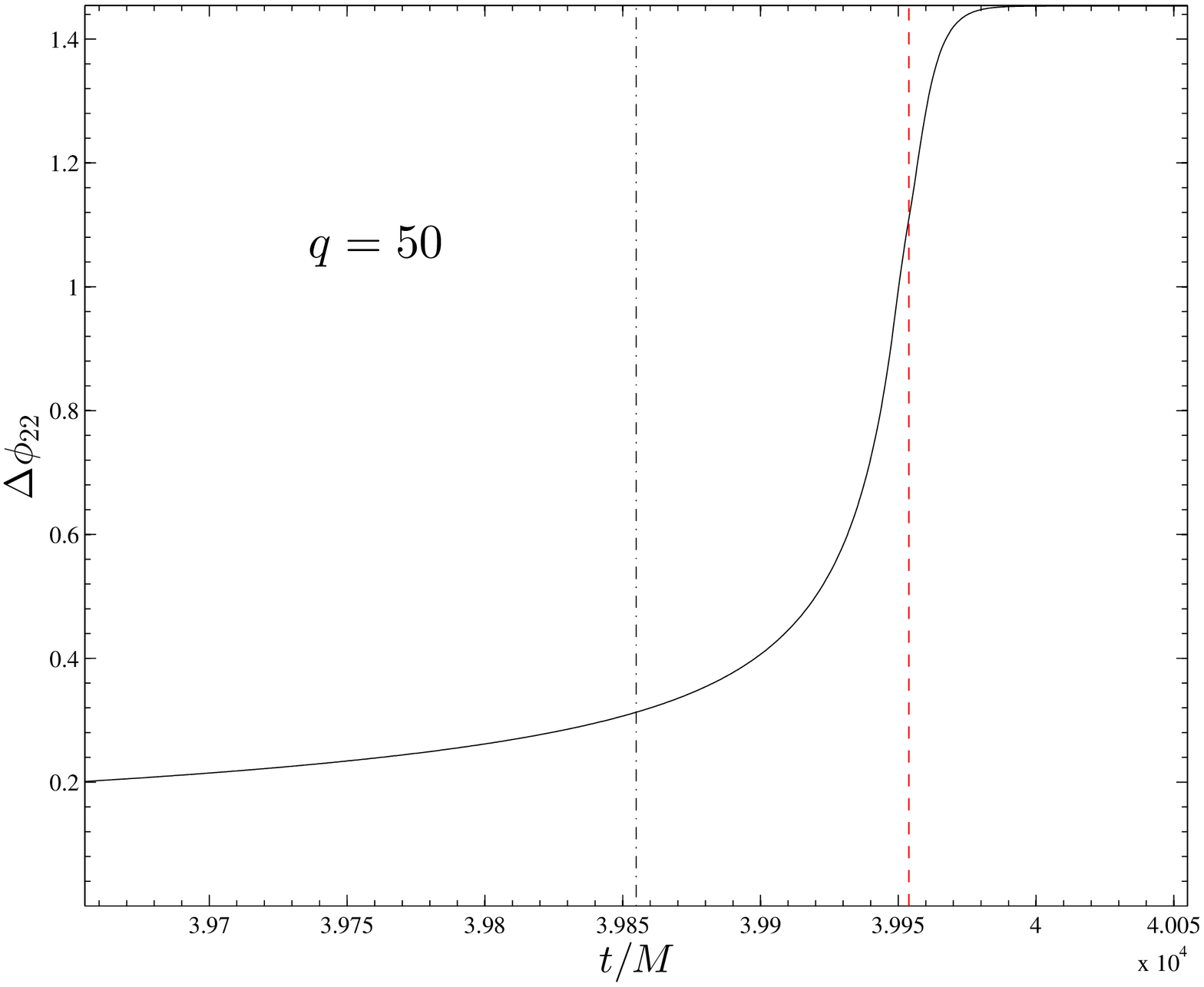}
\includegraphics[width=0.49\textwidth]{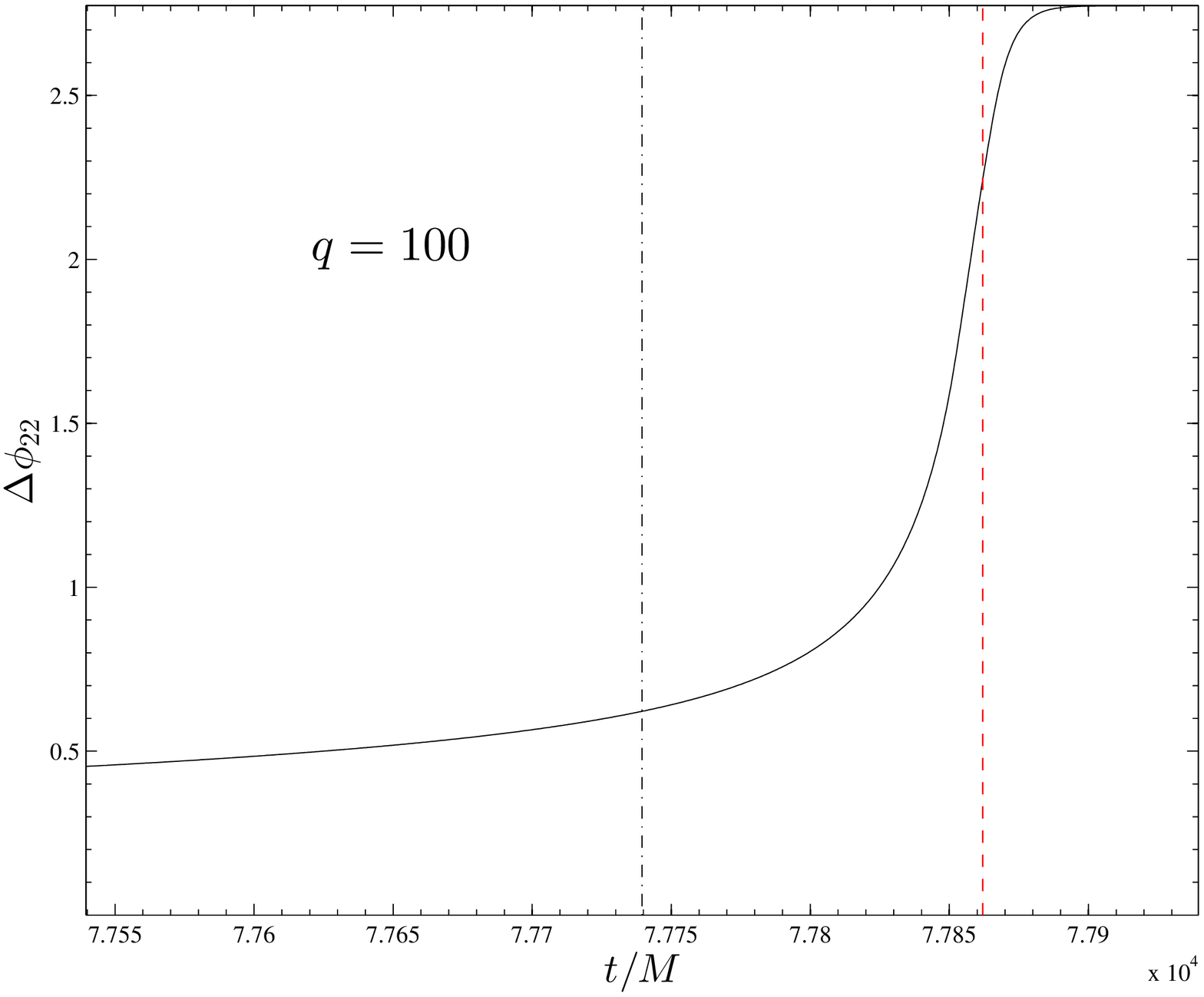}
\caption{\label{fig:dphi_allq} (color online) Accumulated phase difference due to horizon 
absorption for different mass ratios $q$ as obtained from EOB evolutions. The vertical 
lines mark the crossing of the EOB-defined LSO (leftmost line) and of the EOB-defined 
light-ring (rightmost line). For all binaries, the initial separation is $r_0=15$, 
corresponding to $M\omega_{22}^0=0.0344$.}
\end{figure*}

In this section we discuss and quantify the effect of the inclusion of absorbed fluxes, $\F_\varphi^H$, 
in the dynamics on the observable GW (i.e. at infinity) from coalescing nonspinning binaries of 
different mass ratios. We work here only with EOB-generated waveforms.

We focus first on the test-mass limit, $\nu=10^{-3}$, subject to leading-order (in $\nu$) radiation 
reaction, Eq.~\eqref{eq:RR_test_mass} (we neglect then all the higher-order $\nu$-dependent corrections).
The effect of  $\hat{f}^H(v_\varphi^2;\,0)$ on the $\ell=m=2$ phasing is illustrated in Fig.~\ref{fig:dphi_test_mass}.
The initial separation is, as before, $r_0=7$, which yields about 41 orbits up to merger (see Table~\ref{tab:dPhi_EOB}).
The top panel displays the EOB waveform without including horizon absorption (dashed line) together with the 
one where BH absorption is taken into account. The leftmost vertical line marks the LSO crossing, 
while the rightmost vertical line the light-ring crossing.
The visible difference between the two waveforms is made quantitative in the bottom
panel of the figure, where the phase difference is shown.
Here it is $\Delta\phi_{22}=\phi^{H+\scri}-\phi^{\scri}$. 
One sees that the phase difference is~0.1~rad at the LSO and grows up to 1.6~rad at merger.

We turn now to compare a set of GWs from binaries with $q=1,4,10,50,100$ and $1000$, computed using 
the complete EOB dynamics. We run the simulations with and without horizon absorption and we compute the
phase differences. The initial separation for $q=1000$ is $r_0=7$, while for the other mass ratios 
it is $r_0=15$, corresponding to the initial GW frequency $M\omega_{22}^0=0.0344$. The result of this comparison 
is displayed in Fig.~\ref{fig:dphi_allq} and completed quantitatively by Table~\ref{tab:dPhi_EOB}.
In the four panels of Fig.~\ref{fig:dphi_allq}, the vertical lines mark, respectively from the left, 
the adiabatic LSO crossing and the EOB-defined light-ring crossing, i.e. the conventional location of the merger.
First of all, we notice that even in the equal-mass case, where absorption effects are smallest and 
the system has a limited number of cycles, one gets a dephasing of the order of $5\times 10^{-3}$ rad at the EOB merger.
Remarkably, this value is comparable to (or just a little bit smaller than) the uncertainty
on the phase of the most accurate numerical simulations of (equal-mass, nonspinning) coalescing black-hole 
binaries currently available~\cite{Scheel:2008rj,Pollney:2009yz,Reisswig:2009us}.

For higher mass ratios the cumulative effect of a larger horizon absorption
(acting over more GW cycles) produces larger and nonnegligible dephasings. As listed in Table~\ref{tab:dPhi_EOB}, 
mass ratios of $q\sim10$ to $100$ accumulate (respectively) a dephasing of $\Delta\phi_{22}^{\rm  LSO}\sim0.06$ to $0.6$~rad 
at LSO which increases by a factor of 3 near the light ring, $\Delta\phi_{22}^{\rm LR}\sim0.22$ to $2.2$~rad. 
The last two columns in Table~\ref{tab:dPhi_EOB} list the dephasings obtained using the nonresummed (1PN-accurate) radiation reaction. 
Interestingly, using such an expression of the absorbed flux yields
dephasings that are up to $30\%$ smaller ($q=100$) at merger than the EOB prediction, underestimating the actual effect of absorption.

\begin{table*}[t]
\caption{\label{tab:dPhi_EOB}
  Accumulated phase differences due to to horizon absorption for different mass ratios. 
  The data of the first six binaries are obtained from a complete EOB simulations. On the contrary, the dynamics
  of the last binary, shown for comparison, is that of a point-particle driven by leading-order radiation reaction only. 
  For the first five binaries, the initial separation is $r_0=15$, which corresponds to frequency $M\omega_0^{22}\approx 0.0344$,
  while the last two binaries start at $r_0=7$, i.e. $M\omega_0^{22}=0.108$.
  From left to right, the columns report: the mass ratio $q$; the symmetric mass ratio $\nu=q/(1+q)^2$ 
  ($\nu=1/q$ for the last binary); 
  the initial separation; the number of orbits up to merger (EOB-defined light-ring crossing), $N_{\rm orb}$; 
  the dephasing $\Delta\phi_{22}=\phi^{H+\scri}_{22}-\phi^{\scri}_{22}$ accumulated at the (adiabatic) EOB-defined LSO crossing; 
  the corresponding value expressed in GW cycles; the dephasing accumulated at the EOB-defined light-ring crossing; 
  the corresponding value expressed in GW cycles.
  The rightmost two columns show the phase difference accumulated using Taylor-Poisson, nonresummed, 1PN accurate radiation reaction.
  Note that the effect of horizon absorption on the phasing is still nonegligible (for $q\geq 10$) even using this leading order
  approximation to $\hat{\F}_\varphi^H$.}
 \begin{center}
    \begin{ruledtabular}
      \begin{tabular}{ccc|ccccc|cc}
       $q$ & $\nu$ & $r_0$ & $N_{\rm orb}$ &$\Delta\phi_{22}^{\rm LSO}$ [rad] & $\Delta \N^{\rm LSO}$ & $\Delta\phi_{22}^{\rm LR}$ [rad] & $\Delta \N^{\rm LR}$  
 &$\Delta^{\rm 1PN}\phi_{22}^{\rm LSO}$ [rad] & $\Delta^{\rm 1PN}\phi_{22}^{\rm LR}$ [rad]  \\
       \hline 
      1   & 0.250000 & 15 & 15  & 0.003289 &  0.000523 &  0.005475 &  0.000871  & 0.002849 & 0.004547\\
      4   & 0.160000 & 15 & 21  & 0.028725 &  0.004572 &  0.104712 &  0.016665  & 0.012320 & 0.020246\\
     10   & 0.082645 & 15 & 38  & 0.064372 &  0.010245 &  0.220496 &  0.035093  & 0.052834 & 0.199428\\
     50   & 0.019223 & 15 & 153 & 0.312210 &  0.049690 &  1.115319 &  0.177508  & 0.230220 & 0.765105\\
     100  & 0.009803 & 15 & 296 & 0.620662 &  0.098781 &  2.217042 &  0.352853  & 0.458168 & 1.549226\\
     \hline
     1000 & 0.000998 & 7 & 41.2 & 0.129978 &  0.020687 &  1.453992 &  0.231410  & $\dots$ & $\dots$\\
     \hline
     \hline
     1002 & 0.000996 & 7 & 40.9 & 0.129023 &  0.020535 &  1.563971 &  0.248914 & $\dots$ & $\dots$ \\
   \end{tabular}
  \end{ruledtabular}
\end{center}
\end{table*}

Since horizon absorption effects on phasing are relatively large, 
especially for $q>50$, they may be relevant in template modeling for 
large-mass-ratio binaries. In particular, we focus on IMR binaries 
made by a stellar-mass  compact object (SMCO) and an intermediate mass black-hole 
(IMBH), $(M_A,M_B)\sim (1,50-500)\Msun$, that are candidate sources for
Advanced LIGO~\cite{Brown:2006pj}, and for the Einstein 
Telescope (ET)~\cite{Huerta:2010tp}. We perform an indicative calculation 
of the faithfulness ${\cal A}$~\cite{Damour:1997ub} of an EOB template 
without absorption effects in describing a waveform with absorption
effects.
Given two (real) waveforms, say $h_1$ (with horizon absorption) 
and $h_2$ (without horizon absorption) the {\it faithfulness} 
functional~\cite{Damour:1997ub} 
(also denoted with the symbol ${\cal F}$~\cite{Damour:2010zb})
is defined as
\be
\label{eq:amb_fun}
{\cal A}[h_1,h_2]\equiv \max_{\alpha,\tau}\dfrac{(h_1,h_2)}{||h_1|| ||h_2||},
\ee
where the maximization is performed over a relative time $\tau$
and phase shift $\alpha$ between the waveforms, and
\be
\label{eq:wiener}
(h_1,h_2)\equiv 4\Re\int_0^\infty df\dfrac{\tilde{h}_1(f)\tilde{h}_2^*(f)}{S_n(f)},
\ee
defines the Wiener scalar product between the two signals. Here, $S_n(f)$ 
is the one-sided  power spectral density of the detector noise, $\tilde{h}(f)$ 
the (complex) Fourier transform of the signal, and $||h||=(h,h)^{1/2}$ 
the norm associated to the Wiener scalar product. The mass ratios considered 
were $q=10$, $50$, $71.4286$ and $100$, corresponding to total masses 
$M=(10+100)\Msun$, $(10+500)\Msun$, $(1.4+100)\Msun$, and $M=(14+140)\Msun$.
We followed the technical steps of Ref.~\cite{Damour:2010zb} to compute
accurately the Fourier transform of an EOB waveform. 
We computed the faithfulness ${\cal A}$ taking for $S_n$ both the 
{\tt ZERO\_DET\_HIGH\_P} anticipated  sensitivity curve of 
Advanced LIGO~\cite{shoemaker} and that of the planned Einstein 
Telescope (ET)~\cite{Freise:2008dk,Punturo:2010zz,Sathyaprakash:2012jk}.
The numerical values of ${\cal A}$ are listed in Table~\ref{tab:faith}.
Neglecting horizon absorption (for nonspinning binaries) leads to a loss 
of events ($\propto {\cal A}^3$) of, at most, $0.27\%$ (for LIGO) and 
$0.9\%$ for ET. 
These numbers can be considered negligible for practical purposes. 

As a last remark, we argue that absorption fluxes in the nonspinning 
case are negligible also for parameter estimation. We computed a 
simplified {\it effectualness} functional~\cite{Damour:1997ub} by 
considering a maximization over the total binary mass only. For the most
relevant case $q=100$, $M=141.4\Msun$ and the ET sensitivity curve, we
found that $\max_{M} {\cal A}=0.998$. A more detailed study 
of the effectualness would need maximization over every physical 
parameter of the system (e.g., the chirp mass, the symmetric mass
ratio $\nu$ and the spins). Such an extended analysis should be 
performed for the spinning case, where horizon absorption effects 
are more relevant.

\begin{table}[t]
\caption{\label{tab:faith}
  Faithfulness between signals with and without horizon flux for SMCO-IMBH (nonspinning) 
  binaries in the Advanced LIGO and Einstein Telescope sensitivity band. 
  The merger frequency $f_{\rm merger}$ corresponds to the maximum of the 
  EOB waveform modulus $|h_{22}|$}
\begin{center}
\begin{ruledtabular}  
\begin{tabular}{ccccccc}     
    $q$   & $M_A+M_B [\Msun]$  & $f_{\rm merger}$~[Hz] & $\A_{\rm aLIGO}$ & $\A_{\rm ET}$\\
   \hline   
   10     &  $10  + 100$   & 89.16    & 0.9999  & 0.9998   \\
   50     &  $10  + 500$   & 17.92    & 0.9991  & 0.9995  \\
   71.43  &  $1.4 + 100$   & 89.21    & 0.9991  & 0.9983  \\
   100    &  $1.4 + 140$   & 63.63    & 0.9992  & 0.9970  
   \end{tabular}
  \end{ruledtabular}
\end{center}
\end{table}

\section{Conclusions}
\label{sec:conc}

We investigated the importance of horizon absorption effects in
modelling GWs from nonspinning coalescing black hole binaries. 
Considering a recently proposed EOB resummed expression of the absorbed
flux~\cite{Nagar:2011aa}, we verified the EOB expression against
perturbative waveforms from large mass ratio ($q=1000$) binaries
(Sec.~\ref{sec:smallnu}), and explored the effects of absorbed fluxes
on the phasing  considering EOB evolutions for binaries of different
mass ratios $q=1$ to $1000$ (Sec.~\ref{sec:EOB_horizon}).

We tested the accuracy of the analytically resummed horizon 
flux~\cite{Nagar:2011aa}, and in particular of the residual amplitude 
corrections $\rho_{\lm}^H$,  
in the large-mass-ratio, perturbative limit. We compared 
it to the actual horizon flux of angular momentum computed 
solving the Regge-Wheeler-Zerilli equations in the time-domain.  

To improve the accuracy of the perturbative computation, we
employed two hyperboloidal layers~\cite{Zenginoglu:2010cq} 
(horizon-penetrating near the horizon and hyperboloidalÊnear null infinity) 
attached to a compact domain in 
standard Schwarzschild coordinates. This technique, summarized in
Sec~\ref{sec:dlayer}, allows us to include in the computational
domain both null-infinity, $\scri$,  and the horizon, $H$, via 
compactification in the tortoise coordinate. The resulting improvements of 
our perturbative time-domain code combined with high-order finite differencing
lead to such accurate computations of the inspiral and plunge that the late-time
tail of the signal can be calculated very efficiently 
as reported in Appendix \ref{sec:tails}.

We computed the absorbed GW fluxes from the transition 
from inspiral to plunge down to the late inspiral up to merger for the first time. 
We found that the quadrupolar contributions dominate over the subdominant
multipoles accouting for about $98\%$ of the absorbed radiation 
(see bottom panel of Fig.~\ref{fig:insp:jfluxs}).
The $\ell=2$ absorbed angular momentum flux from the perturbative
simulations proved to be consistent at the $1\%$ level with
the analytical expressions proposed in~\cite{Nagar:2011aa}.
Notably, the agreement remains excellent also below the LSO 
crossing and during the plunge. 
The resummation procedure for the flux introduced
in~\cite{Nagar:2011aa} and the numerical determination
of the higher-order PN terms entering the  $\rho_\lm^H$ amplitude 
corrections were crucial to obtain this result. 
The 1PN accurate, Taylor-expanded expression of the horizon flux 
as computed by Taylor and Poisson~\cite{Taylor:2008xy}, underestimates 
horizon absorption by as much as a factor $2$ during the late-inspiral 
and plunge phases.

The absorbed flux of~\cite{Nagar:2011aa} has been used to build an
additional term to the radiation reaction force of the EOB model,
${\cal F}^H_\varphi$, thereby incorporating in the model, in a resummed way,
horizon absorption. By means of EOB simulations we explored its
effect on the phasing of the GW emitted by binaries of different 
mass ratios $q$.  Even in the current nonspinning case, it yields
nonnegligible phase differences for $q>1$. In particular, in the 
mass-ratio range $q=10$ to $100$ (see Table~\ref{tab:dPhi_EOB}),  the 
accumulated phase differences are of the order $0.2$ to $2$~rad up to merger 
for circularized binaries initially at relative separation of $r_0=15$. 
By contrast, the PN-expanded radiation reaction underestimates 
the dephasing by $9\%$ to $48\%$ (depending on $q$). 

Finally, we have performed a preliminary investigation of the impact of 
horizon absorption on the accurate modeling of templates for 
IMR nonspinning binaries made by a SMCO and a IMBH $(M_A,M_B)\sim (1,50-500)\Msun$. 
We found that neglecting ${\cal F}^H_\varphi$ would yield a loss of 
events by $0.27\%$ for Advanced LIGO and by $0.9\%$ for ET. 
These losses are essentially negligible by current accuracy standards.

Horizon absorption effects are more important for spinning binaries. 
It will be necessary to include them in ${\cal F}^H_\varphi$, after 
a suitable resummation procedure, so to study their impact on the 
phasing. Similarly, we expect their influence to be nonnegligible 
on faithfulness and effectualness computations for gravitational 
wave data analysis purposes.

\acknowledgements

We are grateful to S.~Akcay for the numerical data of Fig.~\ref{fig:ciro}, 
and D.~Pollney for giving us access to the NR data 
of~\cite{Damour:2011fu,Damour:2012prep}. We thank T.~Damour for useful
suggestions, and N.K.~Johnson-McDaniel for reading the manuscript.
SB is supported by DFG GrantSFB/Transregio~7 ``Gravitational Wave
Astronomy.''  S.~B. thanks IHES for hospitality and support during the
development of part of this work. 
A.~Z. is supported by the NSF Grant No.~PHY-1068881, 
and by a Sherman Fairchild Foundation grant to Caltech.
Computations were performed on the {\tt MERLIN} cluster at IHES.

\appendix

\section{Late-time tail decay for radial infall and insplunge trajectories}
\label{sec:tails}

In this Appendix we present, for the first time, the  accurate computation 
of the late-time power-law tail of the waveform at $\scri$, generated by a 
particle plunging, both radially and following an inspiralling trajectory, 
into a Schwarzschild black hole.
This result completes the knowledge of the $\scri$-waveform for these events, 
already computed elsewhere~\cite{Bernuzzi:2010xj,Bernuzzi:2011aj}. 

We recall that the gravitational waveform is computed by solving the 
RWZ equations in the time domain for each multipole. 
The $\delta$-function representing the particle is approximated by a 
narrow Gaussian of finite width $\sigma\ll M$, Eq.~\eqref{eq:gauss}. 
The representation of a particle as a Gaussian is a standard method 
when gravitational perturbations are computed using finite-difference, 
time-domain methods.   This representation, however, was considered 
problematic, because time-domain codes gave relatively inaccurate results 
for gravitational fluxes~\cite{Burko:2006ua, Jung:2007zf, Barton:2008eb}. 
Therefore, different prescriptions have been experimented with to improve on 
the representation of the point particle through a Gaussian~\cite{Sundararajan:2007jg, 
Sundararajan:2008zm, Sundararajan:2010sr}.
Nevertheless, the accuracy of time domain codes remained low, especially  
when compared with frequency domain ones. One open problem was the calculation 
of tail decay rates for a particle radially infalling into a Schwarzschild 
black hole~\cite{Jung:2007zf}. 

Recently, a multi-domain hybrid method of finite difference and spectral discretizations 
has been developed to solve this problem~\cite{Chakraborty:2011gx}. With this method, 
and using a large computational domain, the polynomially decaying part of the signal 
could be computed. However, the width of the Gaussian used in~\cite{Chakraborty:2011gx} 
to represent the particle is inadequate for the particle limit. In fact, the ``particle" 
in this study is larger than the Schwarzschild black hole that provides the background.

In this Appendix, we show that the accuracy provided by hyperboloidal layers, combined with high-order finite differencing, allows us to 
calculate the tail decay rates accurately for realistic representations of a point particle in 
Schwarzschild spacetime. We present the decay rates not only for a radially infalling particle, 
but also for an insplunging one. 

As in previous work~\cite{Nagar:2006xv, Bernuzzi:2010ty, Bernuzzi:2010xj, Bernuzzi:2011aj}, 
we approximate the delta distribution that represents the particle at time-dependent location, $R_*(t)$, by a Gaussian
\be\label{eq:gauss}  \delta(r_* - R_*(t) ) \ \to \ \frac{1}{\sqrt{2\pi}\sigma} \exp \left(-\frac{(r_*-R_*(t)^2}{2\sigma^2}\right) . \ee
Our prescription for the standard deviation, $\sigma$, depends on resolution. 
We set $\sigma = 4\,\triangle r_*$, so that the Gaussian is resolved well 
on our finite difference grid. 

Transmitting layers play an essential role in resolving narrow Gaussians because they allow us 
to compute the infinite domain solution in a small grid. This implies that the numerical resolution 
is not wasted in simulating empty space; instead, it can be focussed to where the particle is located. 
As a consequence, we can afford to choose $\triangle r_*$, and therefore the width of the Gaussian 
$\sigma$, very small.

Another advantage of using the layer method is that the implementation of high-order finite differencing 
becomes simpler because there are no boundary conditions to be applied at either end of the domain. 
Note that even when good boundary conditions are available, their discretization and numerical implementation 
may not be straightforward. When no boundary conditions need to be applied, however, using a high order 
finite difference method becomes just a matter of widening the stencils.

Using hyperboloidal layers, we have improved the accuracy of our previous work~\cite{Bernuzzi:2011aj}. 
We use a smaller domain of $[-20,20]$ with interfaces at $R_\pm = \pm12$. Compared to our previous domain 
of $[-50,70]$, this gives us a factor of 3 in efficiency~\footnote{ %
  By construction, reducing domain size does not decrease the time
  step for a given resolution. We did not attempt to find the optimal
  thickness for the layers.}.  
In addition, we use  8th order finite differencing as opposed to 4th order in~\cite{Bernuzzi:2011aj}. 
As a result, we can compute the tail decay rates accurately, as reported below.

\subsection{Radial infall}

The calculation of gravitational perturbations caused by a particle falling radially 
into a non-rotating black hole is a classical problem in relativity~\cite{Zerilli:1971wd,Davis:1971gg}. 
It serves as a good test bed for numerical computations, and there are still relatively 
recent studies on the problem~\cite{Jung:2007zf,Mitsou:2010jv, Chakraborty:2011gx}. 

We solve the radial infall of a particle to demonstrate the accuracy of our infrastructure. 
For a detailed description of the setup, the reader is referred to the
literature~\cite{Lousto:1996sx, Martel:2001yf, Martel:2003jj, Nagar:2005ea}.

\begin{figure}[t]
\center
\includegraphics[width=0.47\textwidth]{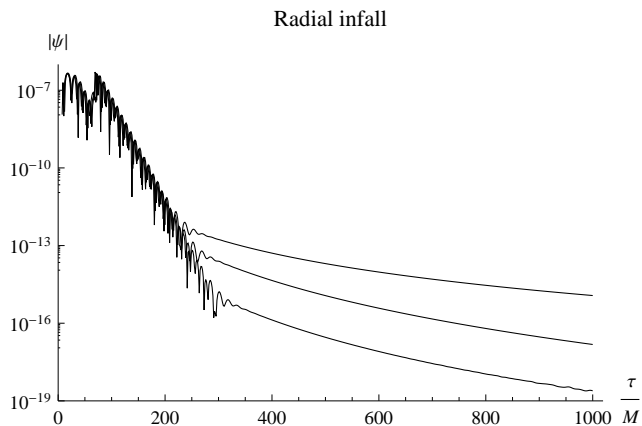}
\caption{\label{fig:infall} The evolution of the field for a radially infalling particle starting at $r_0=7$. 
The plot spans 13 orders of magnitude. Observers are located (from top to bottom) at $\scri$, $30$, and $15$. }
\end{figure}

In Fig.~\ref{fig:infall}, we show the absolute value of the Zerilli function $\psi_{20}$ 
 caused by an infalling particle initially at rest at $r_0=0$ as measured by three observers. The particle is represented 
by the Gaussian~\eqref{eq:gauss} with a full width at half maximum (FWHM)\footnote{The FWHM 
of a normal distribution is given by its standard deviation 
$\sigma$ as $2\sigma \sqrt{2\ln2}$.} of $0.04M$. We use 10,000 grid cells and a 
time stepping factor of $0.75$ for the computation. Note that, differently from 
Refs.~\cite{Lousto:1996sx,Martel:2001yf} we put $\psi_{20}=0$ initially and we
do not solve consistently the Hamiltonian constraint. Since we are interested here
in the late-time behavior of the waveform, this simplifying choice has no influence
on our results.
We see the QNM ringing after the plunge of the particle into the black hole, 
followed by late-time decay. The three curves in the figure correspond to the 
measurements of three observers (from top to bottom): the observer at infinity, 
the finite distance observer at $30M$, and at $15M$. The perturbations are 
computed for about $1000M$ which leads to a drop in the absolute value of 
the perturbation by $13$ orders of magnitude. The polynomially decaying 
signal is reproduced accurately. 

The gain in accuracy is partly a result of the 8th order finite differencing, 
but mostly due to the high resolution we can afford using hyperboloidal 
layers, which allow us not only to compute the perturbations as measured 
by the observer at infinity, but also to follow the signal much longer than is possible with standard methods. 
For example, in Ref.~\cite{Chakraborty:2011gx} the authors 
compute the perturbations until about $600M$ for a Gaussian source that has a 
FWHM of $5-10M$ which is larger than the size of the central black hole, 
and therefore cannot represent a realistic particle\footnote{The representation of the Gaussian in 
\cite{Chakraborty:2011gx} leads to a FWHM of $2\sqrt{\sigma \ln 2}$. The authors present studies with 
$\sigma$ ranging between 10 and 50.}.

\begin{figure}[t]
\center
\includegraphics[width=0.47\textwidth]{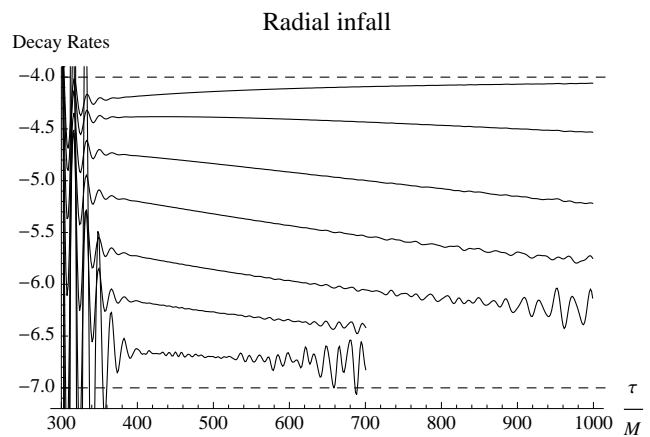}
\caption{\label{fig:tail:infall} The local decay rates for the above evolution. 
The observers are located approximately at (from top to bottom) in units of $M$: 
$\{ \scri, 250, 80, 50, 35, 25, 20 \}$. The dashed lines indicate the theoretically 
expected asymptotic decay rates: $-4$ at $\scri$, and $-7$ at finite distances.}
\end{figure}

We also plot the local decay rates as measured by different far away observers in Fig.~\ref{fig:tail:infall}. 
The local decay rate plot gives a clear image of the accuracy of our computation. We see that the expected 
decay rates are reproduced accurately. The observer at infinity measures a rate of $-4$, whereas the rate 
for finite distance observers approaches $-7$. The intermediate behavior for the decay 
rates for these observers is in accordance with computations of vacuum perturbations~\cite{Zenginoglu:2009ey}. 

The local rates for the observers at $25M$ and $20M$ in Fig.~\ref{fig:tail:infall} have been cut from the 
plot at late times because of large oscillations. The loss of accuracy for these observers is not 
only because of accumulated truncation error, but mostly because the fast decaying signal reaches 
machine precision. If necessary, 
the decay rate calculation can be further improved by using quadruple precision, and possibly higher resolution.

\begin{figure}[t]
\center
\includegraphics[width=0.47\textwidth]{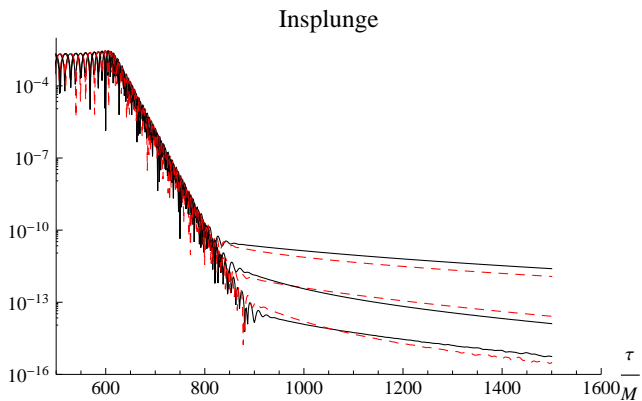}
\caption{\label{fig:insplung} The evolution of the real (solid line) and imaginary (dashed line) of the field for insplunge 
from $r_0=7$. The evolution spans 14 orders of magnitude. Observers are located (in units of $M$, from top to bottom) 
at $\scri$, 35, and 18.}
\end{figure}

\subsection{Insplunge}

The main interest in this paper is the study of particles plunging into the central black hole following a phase 
of quasi-circular inspiral (insplunge). We compute the tail decay rates also for this case. As above, the initial
separation is $r_0=7$.  
In Fig.~\ref{fig:insplung} we show the absolute value of the real part (solid line) and imaginary part (dashed line) of the 
perturbation, again as measured by three observers (from top to bottom): the observer at infinity and the finite distance 
observers at $35M$ and $18M$. The computational parameters are the same as in the radial infall study. 
We see that the field is followed for 14 orders of magnitude, and the evolution is presented until $1500M$ this time. 
The three stages of the evolution (inspiral, ringing, and polynomial decay) are clearly visible. The local decay rates 
show qualitatively the same behavior as in Fig.~\ref{fig:tail:infall} and are therefore not plotted.

\bibliography{refs20121219.bib}

\end{document}